\documentclass[aps,prd,amsmath,amssymb,preprintnumbers,nofootinbib,a4paper,11pt]{revtex4-1}
\pdfoutput=1
\usepackage{amsthm}
\usepackage{graphicx}
	\graphicspath{{./NewFig/}}
\usepackage{url}
\usepackage[bookmarks, pagebackref=false]{hyperref}
\usepackage{color}
	\definecolor{rossoCP3}{cmyk}{0,.88,.77,.40}
		\definecolor{graa}{rgb}{0.8,0.8,0.8}
		\definecolor{blaa}{rgb}{0.2,0.2,0.6}
		\hypersetup{
			colorlinks, 
			bookmarksopen, 
			bookmarksnumbered,
			citecolor=blaa, 		
			linkcolor=rossoCP3,	
			urlcolor=rossoCP3,			
			}
\usepackage{dcolumn}
\usepackage{bm}
\usepackage{bbm}
\usepackage{pxfonts}

\usepackage{epsfig}
\usepackage{placeins}

\usepackage[ margin=5pt, font=small,labelfont=bf, justification=raggedright]{caption}
\usepackage{youngtab}
\usepackage{slashed}
\usepackage{subfig}

\newcommand{\beq}{\begin{eqnarray}}
\newcommand{\eeq}{\end{eqnarray}}

\newcommand{\bmp}{\noindent\begin{minipage}{16cm}}
\newcommand{\emp}{\end{minipage}\vskip 7mm} 

\def\lsim{\mathrel{\rlap{\lower4pt\hbox{\hskip1pt$\sim$}}
    \raise1pt\hbox{$<$}}}                
\def\gsim{\mathrel{\rlap{\lower4pt\hbox{\hskip1pt$\sim$}}
    \raise1pt\hbox{$>$}}}                

\begin{document}
\title{\Large  \color{rossoCP3}  Exceptional and Spinorial Conformal Windows}
\author{Matin Mojaza$^{\color{rossoCP3}{\varheartsuit}}$}\email{mojaza@cp3-origins.net} 
\author{Claudio Pica$^{\color{rossoCP3}{\varheartsuit}}$}\email{pica@cp3-origins.net} 
\author{Thomas A. Ryttov$^{\color{rossoCP3}{\clubsuit}}$}\email{ryttov@physics.harvard.edu} 
\author{Francesco Sannino$^{\color{rossoCP3}{\varheartsuit}}$}\email{sannino@cp3-origins.net} 
 \affiliation{
$^{\color{rossoCP3}{\varheartsuit}}${ \color{rossoCP3}  \rm CP}$^{\color{rossoCP3} \bf 3}${\color{rossoCP3}\rm-Origins} \& the Danish Institute for Advanced Study {\color{rossoCP3} \rm DIAS},\\ 
University of Southern Denmark, Campusvej 55, DK-5230 Odense M, Denmark.
}
\affiliation{$^{\color{rossoCP3}{\clubsuit}}$  \mbox{Center for the Fundamental Laws of Nature - Jefferson Physical Laboratory}\\
\mbox{Harvard University - Cambridge, MA 02138 - USA}} 
\begin{abstract}
We study the conformal window of gauge theories containing fermionic matter fields, { 
where the gauge group is any of the exceptional groups with the fermions transforming according to the fundamental and adjoint representations and the orthogonal groups where the fermions transform according to a spinorial representation}. We investigate the phase diagram using a purely perturbative four loop analysis, the all-orders beta function and the ladder approximation. \\ 
[.1cm]
{\footnotesize  \it Preprint: CP$^3$-Origins-2012-016 \& DIAS-2012-17}
 \end{abstract}

\maketitle

\section{Introduction} 

Over the past few decades strongly interacting theories have continued to pose a considerable challenge for theoretical physicists.  In recent years a thorough and comprehensive study of the phase diagram of non-supersymmetric gauge theories has been undertaken, for a review see \cite{Sannino:2009za}. In brief, the renewed interest is due to the fact that a better knowledge of the nature of the conformal window can lead to the correct theory behind electroweak symmetry breaking. These investigations can also be used for the construction of models of composite dark matter, reviewed in \cite{Sannino:2009za}, and inflation \cite{Channuie:2011rq}. The first series of works were dedicated to the study of $SU(N)$ gauge theories with fermions transforming according to higher dimensional representations of the gauge group  \cite{Sannino:2004qp,Dietrich:2006cm}. In \cite{Sannino:2004qp} it was first realized that only two Dirac fermions were needed in order to be near, or within, the conformal window. Subsequently the study was generalized to non-supersymmetric $SO(N)$ and $Sp(2N)$ gauge theories \cite{Sannino:2009aw} and to supersymmetric gauge theories with superfields belonging to arbitrary representations of the gauge group \cite{Ryttov:2007sr}. The status for the conformal window of chiral gauge theories was summarized and further extended in \cite{Sannino:2009za}.  Except for the supersymmetric case, and the case of the chiral gauge theories, all the investigations were done primarily using the ladder approximation \cite{Appelquist:1988yc,Cohen:1988sq,Miransky:1996pd}. 

However, it was clear that one cannot rely on just one crude approximation and new techniques had to be developed in order to obtain a firmer grip on the phase diagram. Inspired by the work of Seiberg \cite{Seiberg:1994pq} in supersymmetric gauge theories and its use of the NSVZ beta function \cite{Novikov:1983uc} to bound the conformal window an all-orders beta function for fermionic gauge theories was conjectured \cite{Ryttov:2007cx}. Using additional consistency checks it was found that the original beta function had to be slightly corrected in order to accommodate known analytical results \cite{Pica:2010mt}. It is worth emphasizing that the all-orders beta function is shape preserving when approaching the supersymmetric limit and that the beta function is written linearly in the anomalous dimension of the mass similarly to the supersymmetric case. Also a related form of an all-orders beta function was conjectured in \cite{Antipin:2009wr,Antipin:2009dz}. 

One of the key outcomes using the all-orders beta function is its prediction of the anomalous dimension of the mass at a fixed point. In general it was found that the anomalous dimension was somewhat smaller than what was expected from the ladder approximation. Therefore a third method using higher orders perturbation theory was utilized \cite{Ryttov:2010iz,Pica:2010xq}. Both the beta function and the anomalous of the mass were computed to the fourth loop order in the $\overline{MS}$ scheme in \cite{vanRitbergen:1997va,Vermaseren:1997fq} which made it possible to investigate the corrections of higher loop orders in perturbation theory to the anomalous dimension at a fixed point. In general it was found that the anomalous dimension decreases when one includes higher orders signaling the accuracy and { potential} exactness of the all-orders beta function. The study of higher loop orders in supersymmetric theories has also been done to the three loop level in the $\overline{DR}$ scheme \cite{Ryttov:2012qu}. Here the same tendency as in the nonsupersymmetric case with a decreasing anomalous dimension at the fixed point is found. 

Finally we note that also the conformal house of non-supersymmetric gauge theories with fermions transforming according to multiple representations of the gauge group has been of interest \cite{Ryttov:2009yw,Ryttov:2010hs} while the non-trivial consistency checks of the conformal window using dualities have been considered in \cite{Sannino:2009qc,Sannino:2009me,Mojaza:2011rw}.

A quick search through the literature reveals that a considerable amount of the work done so far has been focused primarily on $SU(N)$ gauge theories with fermions transforming according to higher dimensional representations while using a variety of different techniques. Only a few scattered departures away from this direction of research has been carried out. This is certainly an incomplete survey of all possible non-supersymmetric gauge theories. We therefore take the analysis one step further and { exhaust the investigations by studying also theories with exceptional gauge groups and theories with fermions transforming according to spinorial representations}. This is done using a number of different techniques that have all gained their respect in the case of an $SU(N)$ gauge group. They are

\begin{itemize}
\item  The four loop beta function and anomalous dimension of the mass. 
\item The all-orders beta function.
\item The ladder approximation.
\end{itemize}

In the past the use of the exceptional groups and the spinorial representations has found applications in many fields of particle physics including the famous examples of Grand Unified Theories (GUT's) and string theory. Following the famous work on the $SU(5)$ GUT \cite{Georgi:1974sy} where the SM fermions are assigned to the conjugate fundamental and the two-indexed antisymmetric representation other unifying gauge groups were proposed. Specifically the orthogonal group $SO(10)$ was considered where a single generation of SM fermions filled out a complete $16$ dimensional spinorial representation \cite{GeorgiSO(10),Fritzsch:1974nn}. As opposed to the original $SU(5)$ theory the $SO(10)$ theory also allowed the inclusion of a right handed neutrino. Similarly the exceptional group $E_6$ - which has $SO(10)$ as a subgroup - can be used as a unifying gauge group to which all the SM interactions and matter particles can be neatly incorporated \cite{Gursey:1975ki}. For an early review see \cite{Langacker:1980js}. Also in string theory the use of exceptional groups has found its way. Consistency requires that one of the heterotic string theories has $E_8 \times E_8$ as gauge group \cite{Gross:1984dd}. 

From the discussion above it is clear that our analysis might not only provide useful insight into the construction of viable theories able to dynamically break the electroweak theory but it might also shed light on how an eventual Grand Unified Theory should materialize, as well as dark matter and composite inflation.  

Furthermore lattice investigations of exceptional groups \cite{Holland:2003jy,Cossu:2007dk} with the aim to elucidate the relation between chiral symmetry and confinement are already present in the literature granting further support to the present study. 

The paper is organized as follows: In Section \ref{sec:methods} we introduce the various methods and techniques used in our analysis. In Section \ref{sec:exceptional} and \ref{sec:spinorial} we provide the results respectively for the exceptional groups and the spinorial representations. 

\section{Methods and Techniques}\label{sec:methods}

We start by giving a brief description of all the methods that we shall employ to estimate the critical number of flavors above which an IR fixed point is reached. We denote the generators in the representation $r$ of an arbitrary group by $T_r^a,\, a=1\ldots d(G)$. Here $d(r)$ is the
dimension of the representation $r$ and the adjoint
representation is denoted by $G$. The generators are normalized according to
$\text{Tr}\left[T_r^aT_r^b \right] =  T(r) \delta^{ab}$ while the
quadratic Casimir $C_2(r)$ is given by $T_r^aT_r^a =  C_2(r)I$. The
trace normalization factor $T(r)$ and the quadratic Casimir are
connected via $C_2(r) d(r) = T(r) d(G)$. 

Consider a non-abelian gauge group and a set of fermions transforming according to a specific representation of the gauge group. The loss of asymptotic freedom is signaled by the change of sign in the first coefficient of the beta function. The number of flavors for which this occurs is
\begin{eqnarray}
\overline{n}_f &=& \frac{11}{4}  \frac{C_2(G)}{T(r)} \ .
\end{eqnarray}
Just below this  number of flavors the two loop beta function has an infrared fixed point away from the origin. This is the Banks-Zaks perturbative fixed point. As one decreases the number of flavors one expects the fixed point to disappear.

\subsection{Four Loop Analysis}
We first extend the Banks-Zaks perturbative analysis to the maximum known order, to date, in perturbation theory. This will allow to extract relevant information on the perturbative infrared fixed point analysis for the theories investigated here. 

The beta function of a gauge theory with a set of fermions transforming according to an arbitrary representation of the gauge group has been computed to four loop order in the $\overline{MS}$ scheme \cite{vanRitbergen:1997va}. Also the anomalous dimension of the mass is known to this order in the same scheme \cite{Vermaseren:1997fq}. They are given by
\begin{eqnarray}
({8\pi})^{-1}\frac{d\alpha}{d\ln \mu}\equiv\beta(\alpha)   &=& -\beta_0 \left( \frac{\alpha}{4\pi} \right)^2  - \beta_1 \left( \frac{\alpha}{4\pi} \right)^3  - \beta_2 \left( \frac{\alpha}{4\pi} \right)^4   - \beta_3 \left( \frac{\alpha}{4\pi} \right)^5 \ , \\[3mm]
- \frac{d \ln m}{ d\ln \mu^2} \equiv \frac{\gamma(\alpha)}{2} &=& \gamma_0  \frac{\alpha}{4\pi}  + \gamma_1 \left( \frac{\alpha}{4\pi} \right)^2 + \gamma_2 \left( \frac{\alpha}{4\pi} \right)^3 + \gamma_3 \left( \frac{\alpha}{4\pi} \right)^4 \ .
\end{eqnarray}
The various coefficients are given in Appendix \ref{App:Coefficients} and $m$ is the fermion mass. We use the beta function above to determine the  location and type of zeros for the theories investigated here and determine for the infrared fixed point the associated anomalous dimension. We provide also the results for the lower boundary of the conformal window within the four-loop analysis with the caveat that, of course,  higher orders as well as non-perturbative corrections are expected to contribute.  

It is worth mentioning that the four-loop analysis has been useful  for the determination of the anomalous dimensions of $SU(N)$ gauge theories with fermions in various representations of the gauge group. In fact, recent first principle lattice simulations have provided results in reasonable agreement with the theoretical predictions \cite{Giedt:2012rj}.  

The results from this method will be shown together with the other methods in the next section.

\subsection{All-orders Beta Function}

Recently an all-orders beta function for nonsupersymmetric fermonic gauge theories has been proposed \cite{Ryttov:2007cx,Pica:2010mt}. It is given in terms of the anomalous dimension and reads
\begin{eqnarray}
\beta(\alpha) &=& - \left( \frac{\alpha}{4\pi}  \right)^2 \frac{\beta_0 + \frac{\beta_1(\bar{n}_f)}{2\bar{n}_f \gamma_0} n_f \gamma}{1- \frac{\alpha}{4\pi} \frac{\beta_1^{YM}}{\beta_0^{YM}}} \ .
\end{eqnarray}
This beta function is written in terms of the first two universal coefficients as well as the first  universal coefficient of the  anomalous dimension. These are the only scheme independent quantities. $\bar{n}_f$ is the number of flavors above which asymptotic freedom is lost. At the infrared zero of the beta function we have
\begin{eqnarray}
\gamma^\ast &=& \frac{C_2(r)}{2T(r) n_f}  \frac{121C_2(G) - 44T(r) n_f}{7C_2(G) + 11C_2(r)}\ .
\end{eqnarray}
At the infrared fixed point the anomalous dimension is scheme independent and therefore this estimate can be used for any non-supersymmetric vector-like gauge theory with fermions in a given matter representation.

\subsection{Ladder Approximation}

By studying a truncated version of the Dyson-Schwinger equation for the fermion propagator one obtains an estimate for the value of the coupling constant for which the formation of a chiral condensate is triggered and chiral symmetry breaks
\begin{eqnarray}
\alpha_c &=& \frac{\pi}{3C_2(r)} \ .
\end{eqnarray}
To determine when the theory looses conformality traditionally one compares the 
two-loop infrared fixed point value of $\alpha$ with the estimate above. The two-loop value is:
\begin{eqnarray}
\alpha_{IR} &=& - 4\pi \frac{\beta_0}{\beta_1} \ .
\end{eqnarray}
Here the number of flavors is chosen such that the first coefficient of the beta function is larger than zero while the second coefficient is less than zero. Equating the critical value of the coupling above with the value at the IR fixed point yields, what one believes to be, the number of flavors marking the phase boundary of the conformal window
\begin{eqnarray}
n_f &=& \frac{C_2(G)}{T(r)} \frac{17C_2(G)+ 66 C_2(r)}{10C_2(G)+30 C_2(r)} \ .
\end{eqnarray}
In this approach, at this point the anomalous dimension of the mass is expected to be of the order unity.

\section{Conformal Window for the Exceptional Groups}\label{sec:exceptional}
Besides the classical Lie groups there are are also five exceptional ones. These are denoted by $G_2$, $F_4$, $E_6$, $E_7$ and $E_8$. In Table \ref{exceptional} we summarize the various group invariants for the fundamental and adjoint representations. One should note that for  $G_2$, $F_4$, $E_6$ and $E_7$ the fundamental and adjoint representation are distinct while for $E_8$ they coincide. 
The tensor $d^{abcd}$, which appears at four loops, vanishes for all the exceptional groups, see Appendix \ref{App:Coefficients}. Therefore we do not need to compute the fourth order index $I_4$.
All we need is the trace normalization factors, $T(r)$, and the quadratic Casimir, $C_2(r)$, for the fundamental and adjoint representations.

\begin{table}[bt]
\begin{center}
    \begin{tabular}{c||ccccc }
     &\qquad $ G_2 $ &\qquad $ F_4 $ &\qquad $
E_6 $ &\qquad $ E_7 $ &\qquad $ E_8 $   \\
    \hline \hline
    $ C_2(G) $ &\qquad  $ 4b $ &\qquad $ 9b $ &\qquad
     $ 12b $ &\qquad $ 18b $ &\qquad $ 30b $  \\
        $ T(F) $ &\qquad $b$ &\qquad
    $ 3b $
    &\qquad $ 3b $ &\qquad $ 6b $ &\qquad $ - $ \\
    $C_2(F)$ &\qquad  $2b$ &\qquad $6b$ &\qquad $\frac{26}{3}b$ &\qquad $\frac{57}{4}b$ &\qquad $ - $  \\
    $N_F$ &\qquad $ 7 $ &\qquad $ 26 $ &\qquad $ 27 $ &\qquad $ 56 $ &\qquad $ - $ \\
    $ N_G $ &\qquad $ 14 $ &\qquad $ 52 $ &\qquad $ 78 $ &\qquad $ 133 $ &\qquad $ 248 $
    \end{tabular}
\end{center}
\vspace{-5mm}
\caption{Relevant group factors for the exceptional groups.
$b$ is the normalization factor of the Killing form (or equivalently the quadratic Casimir of the adjoint representation) as discussed in \cite{vanRitbergen:1997va, Mojaza:2010cm}.
 The canonical choice for the normalization is $b=1$.}\label{exceptional}
    \end{table}

\begin{figure}[bt]
\centering
\includegraphics[width=0.7\columnwidth]{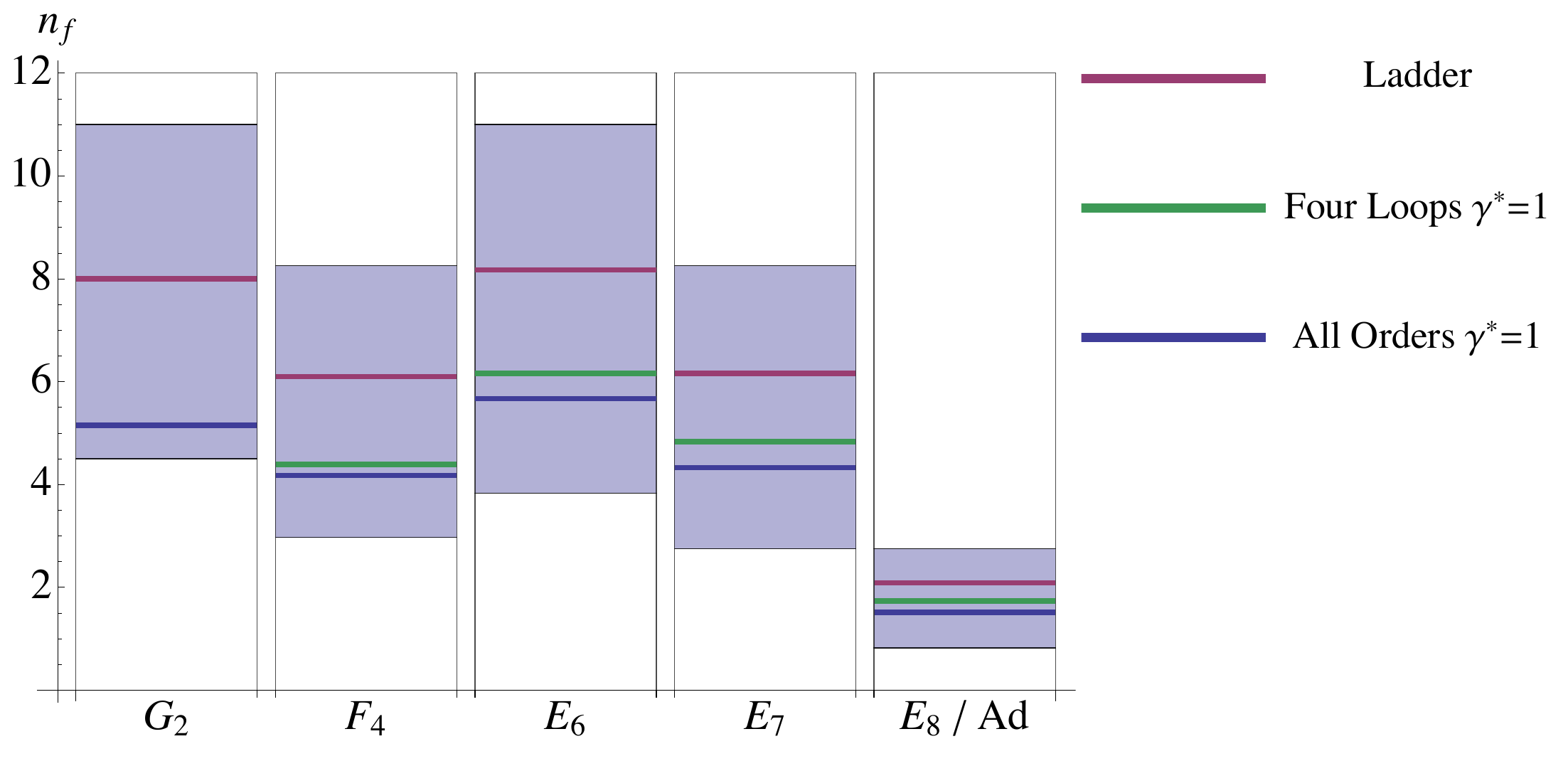}  
 \caption{{ For the exceptional groups we cannot change the number of colors and therefore the lines refer only to the number of flavors per given group. Therefore, we can only construct conformal {\it lines} in the flavor space. Nevertheless, we keep referring to them as conformal \emph{windows}. We plot these lines as bars for the exceptional gauge theories with fermion matter transforming under the fundamental representation.} 
 The upper bound on the window is the asymptotic freedom limit, while the lower border is where the four loop fixed point is lost. 
The legend indicates the estimates from the methods described on the critical number of flavors where large distance conformality is lost.
 For $E_8$ the fundamental and adjoint are identical representations. The results for the theories with fermions in the adjoint representation are almost identical to the one of $E_8$, as can be inspected from Table \ref{Adjoint}.
 }
\label{exceptionally-cool}
\end{figure}

In Tables \ref{Fundamental} and \ref{Adjoint} are provided the various critical number of fermion flavors, as described in the previous section. The upshot of the result is illustrated in Fig.~\ref{exceptionally-cool} .

\begin{table}[b]
\begin{center}
    \begin{tabular}{c||ccccc }
     Fundamental &\qquad $ G_2 $ &\qquad $ F_4 $ &\qquad $
E_6 $ &\qquad $ E_7 $ &\qquad $ E_8 $   \\
    \hline \hline
     Asymptotic freedom  &\qquad  $ 11 $ &\qquad $ 8.25 $ &\qquad
     $ 11 $ &\qquad $ 8.25 $ &\qquad $ - $  \\
     \hline
        Four Loops $\gamma^\ast=1$ &\qquad NA &\qquad
    $ 4.39 $
    &\qquad $6.16  $ &\qquad $ 4.83 $ &\qquad $ - $ \\
    All orders $\gamma^\ast=1$ &\qquad  $ 5.15 $ &\qquad  $ 4.17 $ &\qquad  $ 5.67 $ &\qquad  $ 4.33 $ &\qquad  $ - $ \\
    Ladder &\qquad  $ 8 $ &\qquad  $ 6.1 $ &\qquad  $ 8.17 $ &\qquad  $ 6.16 $ &\qquad  $ - $  \\ \hline
         Loss of 4-loop FP & \qquad $4.50$ & \qquad $2.97$ & \qquad $3.83$ & \qquad $2.75$ & \qquad $-$
    \end{tabular}
  \end{center}
  \vspace{-5mm}
\caption{Table of various critical number of flavors when the fermions are in the fundamental representation. The first list yields the value for which asymptotic freedom is lost. The next three correspond to the critical number of flavors above which an IR fixed point is reached according to the i) four loop beta function and anomalous dimension, ii) the all-orders beta function and iii) the ladder approximation. For all three methods the critical number of flavors has been determined by, or corresponds to, the value of the anomalous dimension at the fixed point being one, $\gamma^*=1$. The last list of critical number of flavors marks the loss of the IR fixed point in the four loop beta function. { NA implies that for that specific theory the four loop anomalous dimension at the fixed point never reaches the value one (see Fig. \ref{GammaG2Fund}). } }\label{Fundamental}
    \end{table}

\begin{table}
\begin{center}
    \begin{tabular}{c||ccccc }
     Adjoint &\qquad $ G_2 $ &\qquad $ F_4 $ &\qquad $
E_6 $ &\qquad $ E_7 $ &\qquad $ E_8 $   \\
    \hline \hline
     Asymptotic freedom  &\qquad  $ 2.75 $ &\qquad $ 2.75 $ &\qquad
     $ 2.75 $ &\qquad $ 2.75 $ &\qquad $ 2.75 $  \\
     \hline
        Four Loops  $\gamma^\ast=1$ &\qquad $1.70$ &\qquad
    $ 1.72 $
    &\qquad $ 1.73  $ &\qquad $ 1.73 $ &\qquad $ 1.73 $ \\
    All orders  $\gamma^\ast=1$ &\qquad  $ 1.51 $ &\qquad  $ 1.51 $ &\qquad  $ 1.51 $ &\qquad  $ 1.51 $ &\qquad  $ 1.51 $ \\
    Ladder &\qquad  $ 2.08 $ &\qquad  $ 2.08 $ &\qquad  $ 2.08 $ &\qquad  $ 2.08 $ &\qquad  $ 2.08 $  \\ \hline
           Loss of 4-loop FP & \qquad $0.75$ &  \qquad $0.80$ & \qquad $0.80$ & \qquad $0.81$ & \qquad $0.82$
    \end{tabular}
\end{center}
\vspace{-5mm}
\caption{Table of various critical number of flavors, as described in Table \ref{Fundamental}, for fermions in the adjoint representation.
} \label{Adjoint}
    \end{table}

In
Appendix \ref{Zerology} is provided the four loop zerology, i.e. the full structure of zeros in the four loop beta functions, for the exceptional groups.

The anomalous dimension at the perturbative fixed point as a function of the number of flavors is given in Appendix \ref{Gamma} .

\section{Spinorial Representations}\label{sec:spinorial}
Below we summarize the group invariants for the spinorial (for short, spin) representations of $SO(N)$. For odd $N$ the spin representation is real while for even $N$ the spin representation is complex (chiral). 

\begin{table}
\begin{center}
    \begin{tabular}{c||ccc }
    $SO(N)$ Representation & $ \quad T(r) /b $ & $\quad C_2(r) /b $ & $\quad
d $  \\
    \hline \hline
    $ S_{\rm real}$ (Odd $N$)  & \quad $ 2^{\frac{N-7}{2}} $ & $\quad \frac{N(N-1)}{16}$ &\quad
     $2^{\frac{N-1}{2}}$  \\
         $ S_{\rm chiral}$ (Even $N$) & \quad $ 2^{\frac{N-8}{2}}$ & $\quad \frac{N(N-1)}{16}$ &\quad
     $2^{\frac{N-2}{2}}$  \\
        $ G $ & $\quad N-2$ &
    $\quad N-2$
    &\quad$\frac{N(N-1)}2{}$  \\
    \end{tabular}
\end{center}
\vspace{-5mm}
\caption{Relevant group factors for the spinorial representations of the $SO(N)$ group. As described in Table \ref{exceptional}, the 
canonical choice is  $b = 1$.
}\label{spin}
    \end{table}

Using the expressions in Table \ref{spin} it is straightforward to compute the critical number of flavors where asymptotic freedom is lost:
\begin{align}
\bar{n}_f = \frac{11}{4} \left( N-2 \right) 2^{-\frac{N}{2}} \kappa(S) \ ,
\end{align}
where $\kappa(S)$ is the spinor representation dependent factor.
For the real spinor representation of $SO(N)$, with $N$ odd it reads:
\begin{align}
\kappa({S_{\rm real}}) = 2^{7/2} = 8 \sqrt{2} \ .
\end{align}
For the chiral spinor representation with $N$ even it instead reads:
\begin{align}
\kappa(S_{\rm chiral}) = 2^{4} = 16 \ .
\end{align}

We provide also the explicit expression for the index $I_4$ of the spin representations as defined in \cite{vanRitbergen:1997va}. This also appeared in the Appendix of \cite{Mojaza:2010cm}. For the real spin representation of $SO(N)$, with $N\geq 5 $ odd we find:
\begin{align}
I_4 (S_{\rm real}) = -2^{\frac{N-9}{2}} b^2 \ ,
\end{align}
while, for the chiral spin representation of $SO(N)$, with $N$ even and $ \geq 6$:
\begin{align}
I_4 (S_{\rm chiral}) = -2^{\frac{N-10}{2}} b^2 \ ,
\end{align}
where in both cases $b$ is the normalization factor of the Killing form (or equivalently the quadratic Casimir of the adjoint representation) as discussed in \cite{vanRitbergen:1997va, Mojaza:2010cm}. The canonical choice is $b = 1$.
In the case of $SO(3)$ and $SO(4)$ the index $I_4$ simply vanishes.

\begin{figure}[tb]
\centering
\includegraphics[width=0.49\columnwidth]{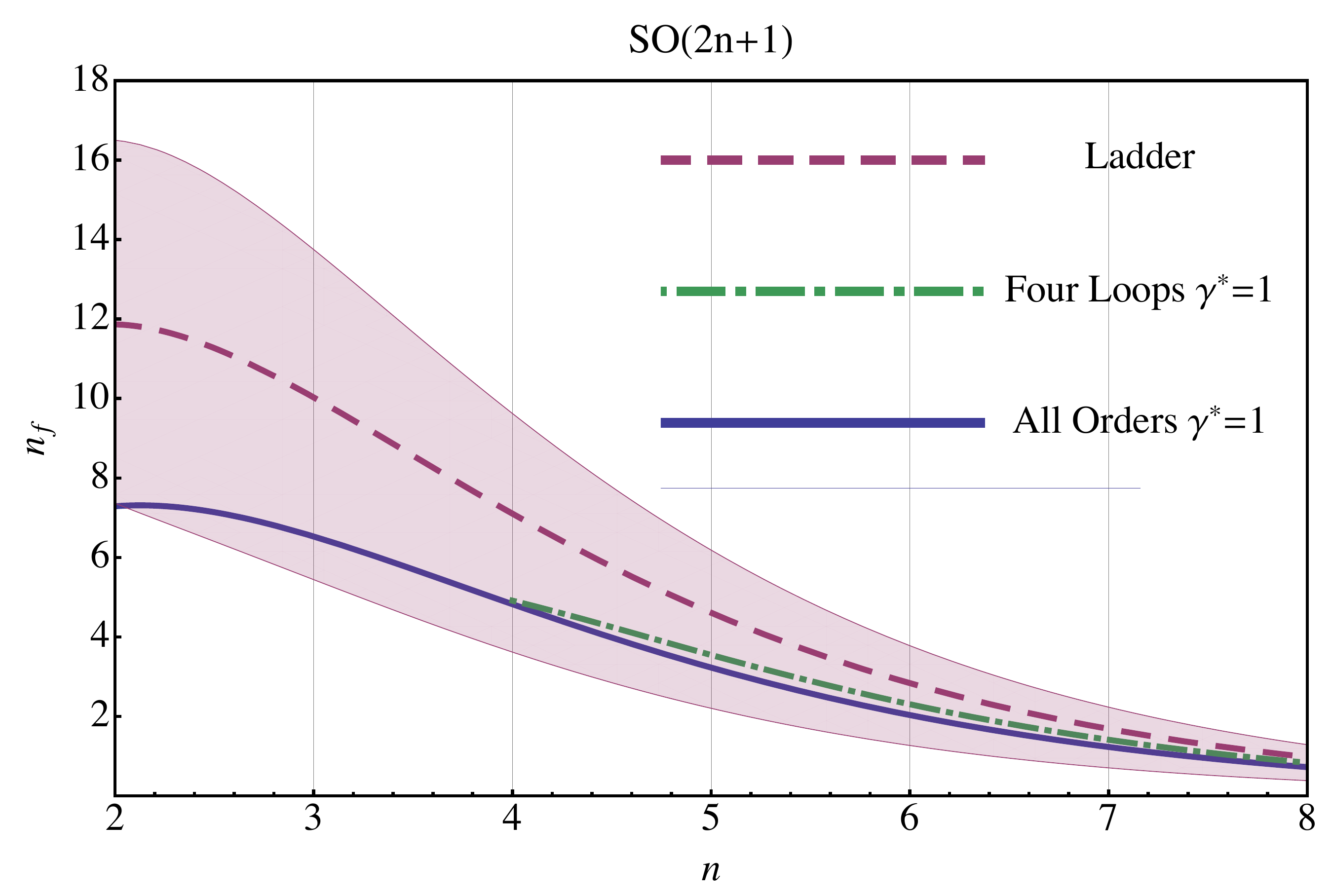} \includegraphics[width=0.49\columnwidth]{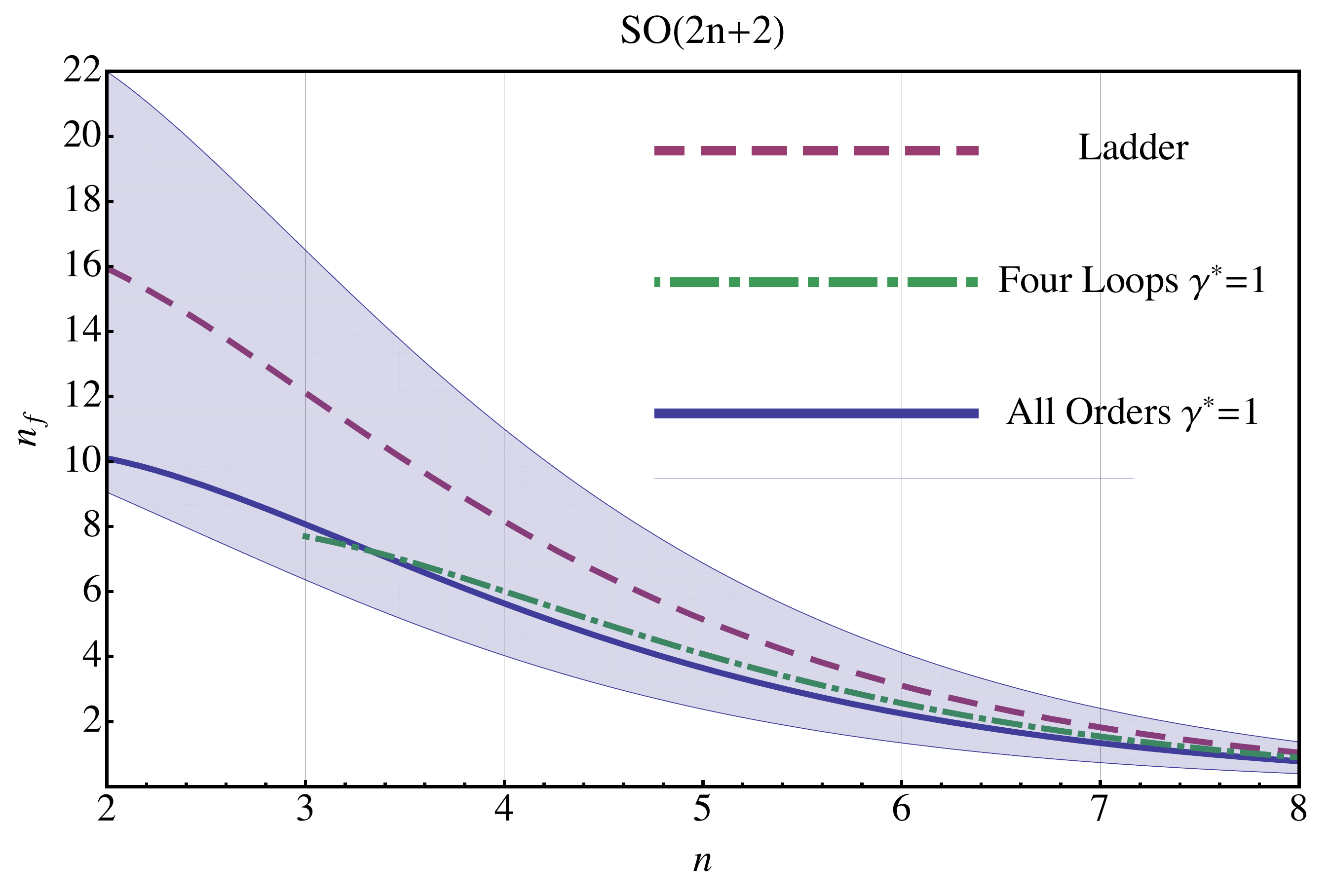} 
 \caption{Conformal window of $SO(N)$ gauge theories with fermion matter transforming under the spinorial representation of the gauge group. The upper bound is the asymptotic freedom limit, while the lower border of the window is where the four loop fixed point is lost. 
 The various estimates of the critical number of flavors where long distance conformality is lost, are shown as given by the legend. All estimates on the critical number of flavors has been determined by, or corresponds to, the value of the anomalous dimension being one, $\gamma^*=1$.  Note that the four loops $\gamma^* =1$ estimate is not applicable for $SO(N<8)$. The all orders and four loop results are in surprisingly good agreement.}
\label{CWSON}
\end{figure}

In Figure \ref{CWSON} is plotted the conformal window of $SO(N)$ gauge theories with fermion matter transforming under the spinorial representation. In the same figures are also shown the various critical values of the number of flavors. The precise numerical values of these are provided in Table \ref{Spinorial}.

\begin{table}[b]
\[
\begin{array}{c ||ccccccccccccccc}
N &  
5 & 6 & 7 & 8 & 9 & 10 & 11 & 12 & 13 & 14 & 15 & 16 & 17 & 18 & 19 \\ \hline \hline
 \text{Asymptotic Freedom} & 
  16.5 & 22. & 13.75 & 16.5 & 9.62 & 11. & 6.19 & 6.88 & 3.78 & 4.12 & 2.23 & 2.41 & 1.29 & 1.38 & 0.73 \\ \hline
\text{Four Loops } \gamma^*=1 & 
\text{NA} & \text{NA} & \text{NA} & 7.7 & 4.91 & 6.01 & 3.55 & 4.08 & 2.31 & 2.56 & 1.41 & 1.54 & 0.83 & 0.9 & 0.48 \\
\text{All Orders } \gamma^*=1 & 
  7.29 & 10.1 & 6.53 & 8.07 & 4.82 & 5.63 & 3.23 & 3.65 & 2.03 & 2.25 & 1.23 & 1.34 & 0.72 & 0.78 & 0.42 \\
 \text{Ladder} &
 11.87 & 15.94 & 10.03 & 12.11 & 7.1 & 8.16 & 4.61 & 5.14 & 2.84 & 3.1 & 1.69 & 1.82 & 0.98 & 1.05 & 0.56 \\ \hline
 \text{Loss of 4-loop FP} &
 7.35 & 9.04 & 5.44 & 6.35 & 3.61 & 4.03 & 2.19 & 2.38 & 1.25 & 1.34 & 0.7 & 0.76 & 0.38 & 0.41 & 0.2
\end{array}
\]
\caption{Table of various critical number of flavors, as described in Table \ref{Fundamental}, for fermions in the spinorial representation of $SO(N)$.
} \label{Spinorial}
    \end{table}

\begin{figure*}[!h]
	\begin{center}
	\subfloat[$SO(5)$. Same topology for $N \leq 5 \wedge N\geq 42$]{\label{SO5} 
	\includegraphics[width=0.49\textwidth]{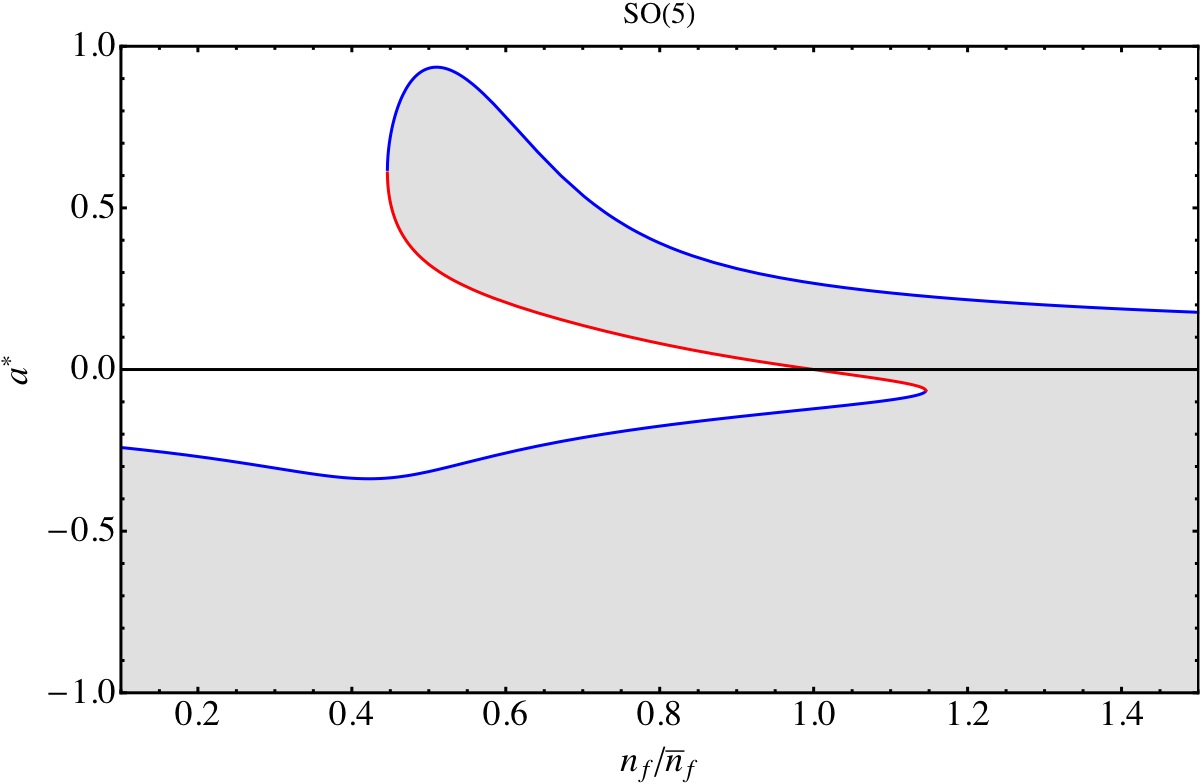}
	}
	%
	\subfloat[$SO(10)$. Same topology for $6 \leq N \leq 12 \wedge N=40,41$]{\label{SO10} 
	\includegraphics[width=0.49\textwidth]{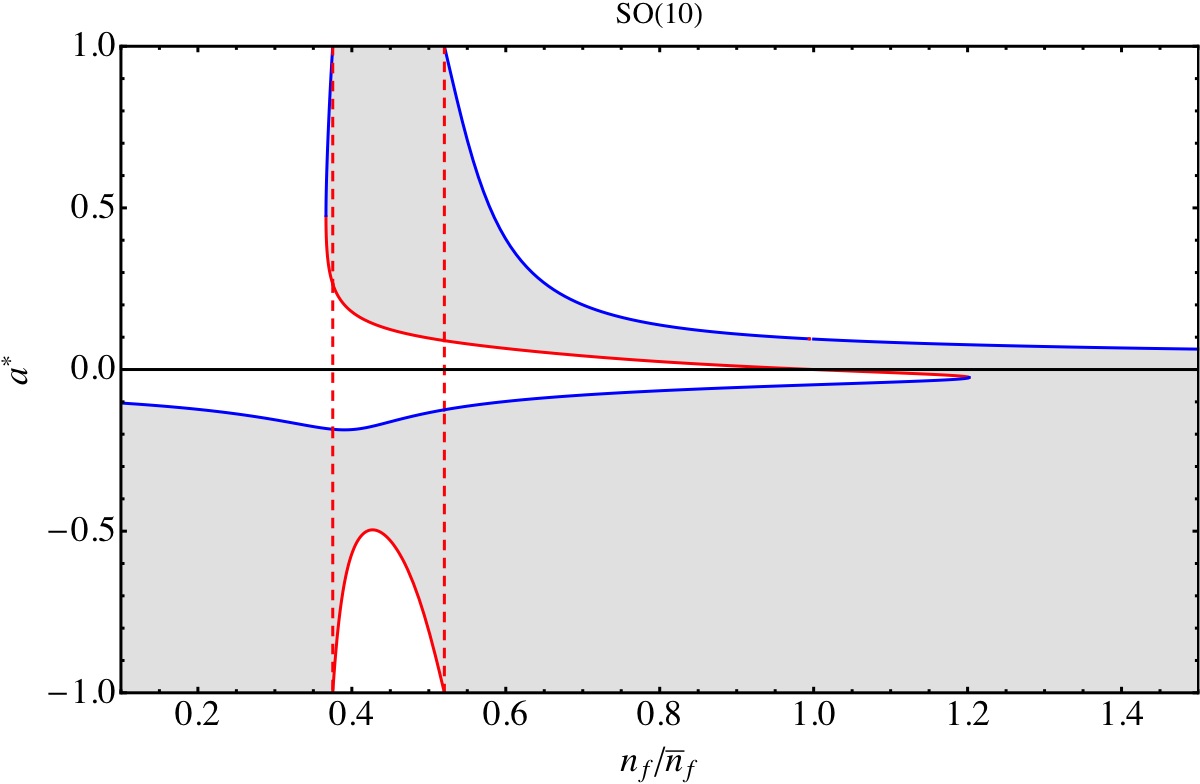}
	}\\
	\subfloat[$SO(13)$. Same topology for $13 \leq N \leq 19 \wedge 30 \leq N \leq 39$]{\label{So13} 
	\includegraphics[width=0.49\textwidth]{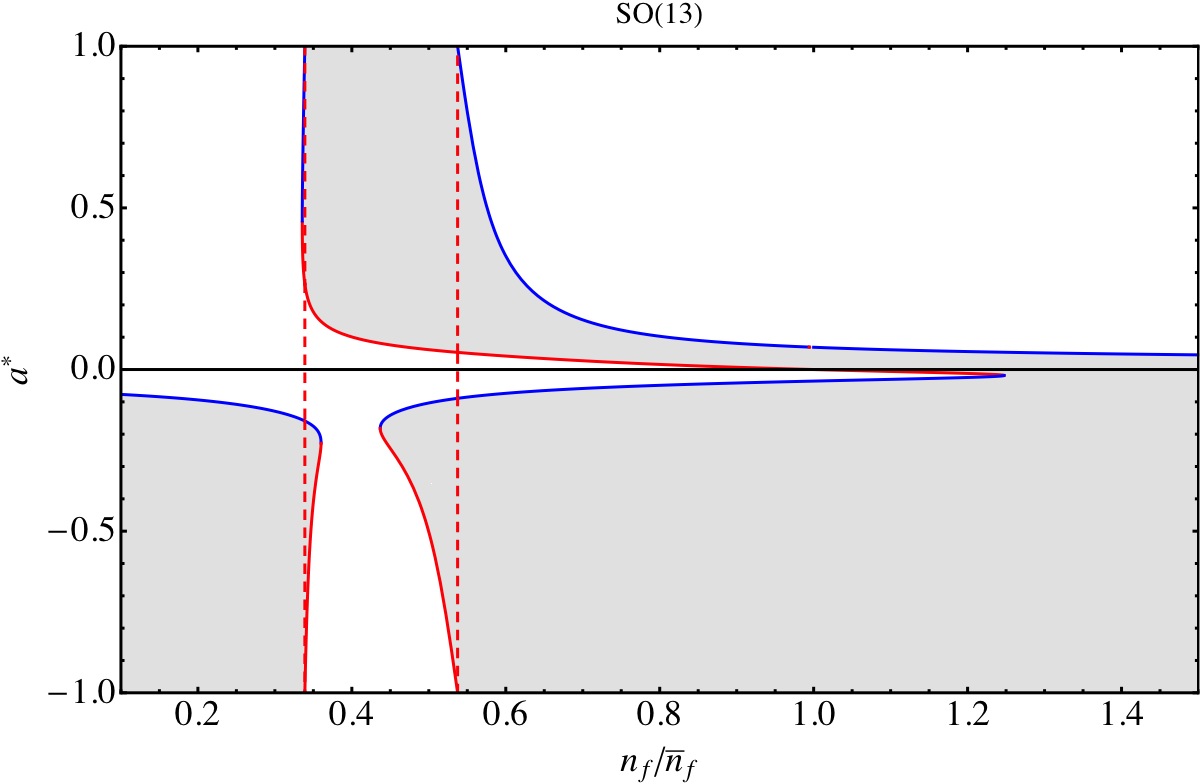}
	}
	%
	\subfloat[$SO(23)$. Same topology for $20 \leq N \leq 29$]{\label{SO23} 
	\includegraphics[width=0.49\textwidth]{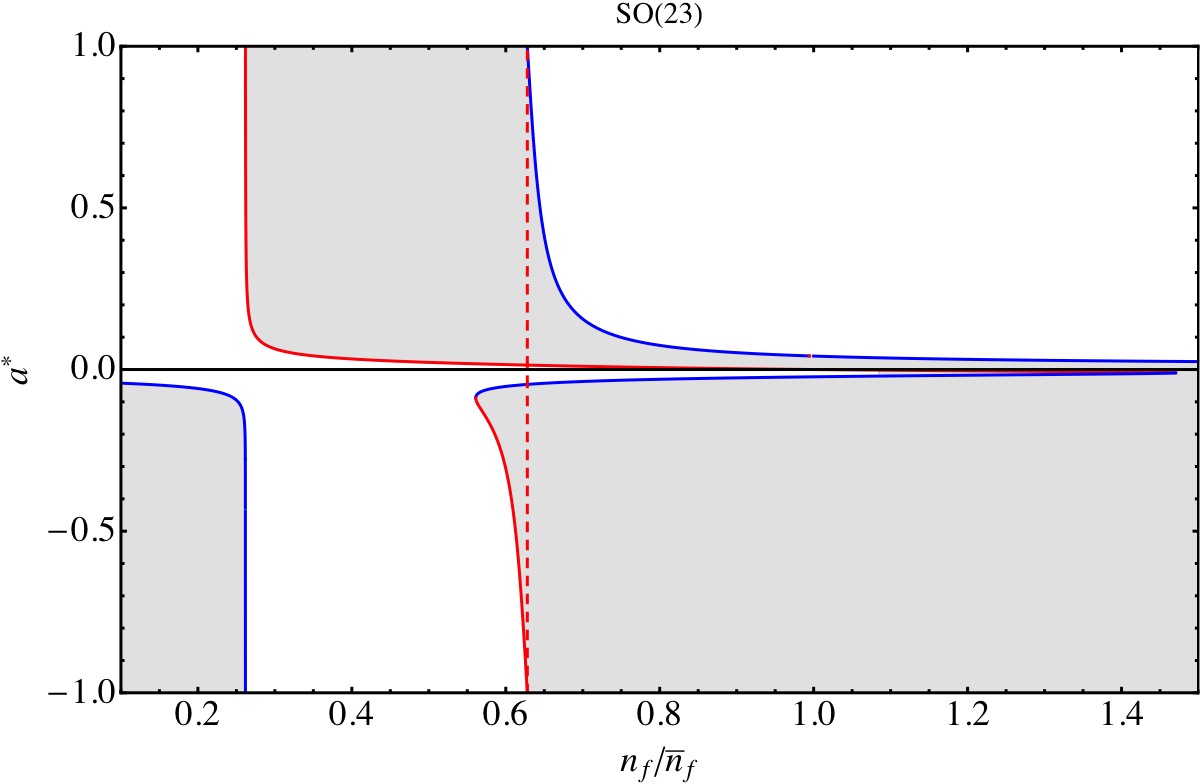}
	}
	\caption{The four distinct \emph{zerologies} of $SO(N)$ gauge theories, as found from the four loops beta functions. The particular groups shown are $SO(5), SO(10), SO(13)$ and $SO(23)$. Red curves correspond to IR fixed point solutions for the given number of flavors, while blue curves correspond to UV fixed point solutions.}
	\label{ZerosSO}
	\end{center}
\end{figure*}

The zerology of the spinorial representations can be classified into four distinct topological structures. In Figure \ref{ZerosSO} four representative plots are shown for each topological structure with classification given for all $SO(N)$ groups.

In Appendix \ref{Gamma} is shown the anomalous dimension at the perturbative fixed point order-by-order in perturbation theory for some of the $SO(N)$ groups.

\section{Conclusions and Outlook}
This work completes the analytic investigation of the conformal window for any asymptotically free gauge group featuring fermionic matter. We discovered that every gauge group has a nontrivial phase diagram relevant when considering various extensions of the standard model. 

To draw the phase diagrams we have used vastly different analytical approaches. Remarkably we have found an extremely good agreement for the prediction of the lower boundary of the conformal window as well as the anomalous dimension of the fermion mass operator using these different methods. We can therefore argue for the presence of an underlying universal structure associated to the phase diagram of any non-supersymmetric gauge theory featuring fermionic matter. The universal structure embodies the existence of a conformal window for a finite number of flavors. The window, however, for the exceptional groups shrinks to a conformal line in the number of flavors. 

What remains to be understood is whether the transition from the conformal to non-conformal phase is of walking \cite{Holdom:1981rm,Holdom:1984sk,Yamawaki:1985zg,Appelquist:1986an} or jumping type \cite{Sannino:2012wy}.  The Dyson-Schwinger analysis, in its most rudimentary incarnation, is unable to address this issue. 

One potentially interesting avenue to explore is the extension of these theories to include other kind of matter fields, even bosonic in nature \cite{Antipin:2011ny,Grinstein:2011dq,Antipin:2011aa,Antipin:2012kc}.  This extensions may lead to new interesting examples of calculable conformal and walking windows \cite{Antipin:2012kc} or could be used to elucidate the spectral dynamics when conformality is lost \cite{Grinstein:2011dq,Antipin:2011aa}. Furthermore these extensions could be of more immediate use for grand unified extensions of the standard model. 
  
\appendix

\section{Coefficients of the Beta Function and the Anomalous Dimension}\label{App:Coefficients}

The four loop beta function coefficients are \cite{vanRitbergen:1997va}:
\label{beta-a}
\begin{widetext}
\begin{eqnarray} 
\label{eq:beta3}
\beta_{0} & = &  \frac{11}{3} C_2(G) - \frac{4}{3} T(r) n_f  \ ,
  \\
 \beta_{1} & = &
 \frac{34}{3}C_2(G)^2 - 4 C_2(r) T(r) n_f -\frac{20}{3} C_2(G) T(r) n_f  \ , \\
 \beta_{2} & = &  \frac{2857}{54} C_2(G)^3
 +2 C_2(r)^2 T(r) n_f - \frac{205}{9} C_2(r) C_2(G) T(r) n_f  - \frac{1415}{27} C_2(G)^2 T(r) n_f   \\ & &
 + \frac{44}{9} C_2(r) T(r)^2 n_f^2
  + \frac{158}{27} C_2(G) T(r)^2 n_f^2  \ ,
\nonumber \\
\beta_{3} & = &
    C_2(G)^4 \left( \frac{150653}{486} - \frac{44}{9} \zeta_3 \right)   
    +  C_2(G)^3 T(r) n_f  
      \left(  - \frac{39143}{81} + \frac{136}{3} \zeta_3 \right)   \\ & &
    + C_2(G)^2 C_2(r) T(r) n_f 
\left( \frac{7073}{243} - \frac{656}{9} \zeta_3 \right) 
   +     C_2(G) C_2(r)^2 T(r) n_f 
      \left(  - \frac{4204}{27} + \frac{352}{9} \zeta_3 \right)   \nonumber \\ & &
   + 46 C_2(r)^3 T(r) n_f 
   +  C_2(G)^2 T(r)^2 n_f^2 
      \left( \frac{7930}{81} + \frac{224}{9} \zeta_3 \right)
    +  C_2(r)^2 T(r)^2 n_f^2 
      \left( \frac{1352}{27} - \frac{704}{9} \zeta_3 \right)\nonumber \\ & &
    +  C_2(G) C_2(r) T(r)^2 n_f^2 
      \left( \frac{17152}{243} + \frac{448}{9} \zeta_3 \right)
  + \frac{424}{243} C_2(G) T(r)^3 n_f^3   
   + \frac{1232}{243} C_2(r) T(r)^3 n_f^3  \nonumber \\ & & 
       +  \frac{d_G^{a b c d}d_G^{a b c d}}{N_G }  
             \left(  - \frac{80}{9} + \frac{704}{3} \zeta_3 \right)
       + n_f \frac{d_G^{a b c d}d_r^{a b c d}}{N_G }  
            \left(   \frac{512}{9} - \frac{1664}{3} \zeta_3 \right)
       + n_f^2 \frac{d_r^{a b c d}d_r^{a b c d}}{N_G }   
            \left(  - \frac{704}{9} + \frac{512}{3} \zeta_3 \right)  \ .
 \label{mainbeta} \nonumber 
\end{eqnarray} 
\end{widetext}

The coefficients of the anomalous dimension to four loops are \cite{Vermaseren:1997fq}:
 \renewcommand{\arraystretch}{1.3}
\begin{widetext}
\begin{eqnarray} 
 \label{maingamma}
\gamma_{0} & = &
 3 C_2(r) \ ,
 \\
 \gamma_{1} & = &
 \frac{3}{2}C_2(r)^2+\frac{97}{6} C_2(r) C_2(G) -\frac{10}{3} C_2(r) T(r) n_f \ ,
 \\
 \gamma_{2} & = &
  \frac{129}{2} C_2(r)^3 - \frac{129}{4}C_2(r)^2 C_2(G)
 + \frac{11413}{108}C_2(r) C_2(G)^2  
 +C_2(r)^2 T(r) n_f (-46+48\zeta_3) 
 \nonumber \\ & &
+C_2(r) C_2(G) T(r) n_f \left( -\frac{556}{27}-48\zeta_3 \right)  
- \frac{140}{27} C_2(r) T(r)^2 n_f^2 \ ,
 \\
\gamma_{3} & = & 
  C_2(r)^4 \left(-\frac{1261}{8} - 336\zeta_3 \right)
   + C_2(r)^3 C_2(G) \left( \frac{ 15349}{12} + 316 \zeta_3 \right)
 \nonumber \\ & & 
   + C_2(r)^2 C_2(G)^2 \left(-\frac{ 34045}{36} - 152 \zeta_3 + 440\zeta_5 \right)
   + C_2(r) C_2(G)^3 \left( \frac{70055}{72} + \frac{1418}{9} \zeta_3
                      - 440 \zeta_5 \right)
 \nonumber \\ & &
  + C_2(r)^3 T(r) n_f \left( -\frac{280}{3} + 552 \zeta_3 - 480 \zeta_5 \right)
  + C_2(r)^2 C_2(G) T(r) n_f \left(- \frac{8819}{27} + 368 \zeta_3 
                            - 264 \zeta_4 + 80 \zeta_5 \right)
 \nonumber \\ & &
  + C_2(r) C_2(G)^2 T(r) n_f \left(- \frac{65459}{162} 
                  - \frac{2684}{3} \zeta_3 + 264 \zeta_4
                    + 400 \zeta_5 \right)
  + C_2(r)^2 T(r)^2 n_f^2 \left( \frac{304}{27} - 160 \zeta_3 
                            + 96 \zeta_4 \right)
  \nonumber \\ & &
  + C_2(r) C_2(G) T(r)^2 n_f^2 \left( \frac{1342}{81} 
                             + 160 \zeta_3 - 96 \zeta_4 \right) 
  + C_2(r) T(r)^3 n_f^3 \left(- \frac{664}{81} + \frac{128}{9} \zeta_3 \right)
 \nonumber  \\ & & 
  + \frac{d_r^{a b c d}d_G^{a b c d}}{N_r}   
            \left(- 32 + 240 \zeta_3  \right)
  +  n_f \frac{d_r^{a b c d}d_r^{a b c d}}{N_r}   
            \left( 64 - 480 \zeta_3  \right)  \ . 
 \end{eqnarray}
\end{widetext}

In the above expressions $\zeta_x$ is the Riemann zeta-function evaluated at $x$, $T_r^a$ with $a=1,\ldots, N_r$ are the generators for a generic representation $r$ with dimension $N_r$.  The generators are normalized via tr$(T^a_r T^b_r) = T(r) \delta^{a b}$ and the quadratic Casimirs are  $[T^a_r T^a_r]_{ij} = C_2(r) \delta_{ij}$. The representation $r=G$ refers to the adjoint representation. The number of fermions is indicated by $n_f$.

The symbol $d_r^{abcd}$ 
is the following fully 
symmetrical tensors:
\begin{eqnarray}
 d_r^{a b c d} & = & \frac{1}{6 } {\rm Tr }  \left[
   T^a_r T^b_r T^c_r T^d_r
 + T^a_r T^b_r T^d_r T^c_r 
 + T^a_r T^c_r T^b_r T^d_r  \right. \nonumber \\
 & & \left. \hspace{4mm}
 + T^a_r T^c_r T^d_r T^b_r 
 + T^a_r T^d_r T^b_r T^c_r 
 + T^a_r T^d_r T^c_r T^b_r  
  \right] \hspace{1mm}
\end{eqnarray}
The contractions can be written purely in terms of group invariants:
\begin{align}
d_r^{abcd}d_{r^\prime}^{abcd} = I_4(r)I_4(r^\prime) d^{abcd}d^{abcd} + \frac{3 N_G}{N_G+2} T(r)T(r^\prime)\left(C_2(r)-\frac{1}{6}C_2(G)\right)\left(C_2(r^\prime)-\frac{1}{6}C_2(G)\right) \ .
\end{align}
The expressions for the relevant group invariants are given in the main text.
As mentioned there, $I_4(r)$ vanished for all exceptional groups and
for $SO(3)$ and $SO(4)$.
The tensor $d^{abcd}$ is representation independent, but not group independent, and the value of its contraction for the groups $SU(N)$, $SO(N)$ and $Sp(N)$ was given in \cite{vanRitbergen:1997va}. Here it is only relevant to quote the $SO(N)$ case:
\begin{align}
d^{abcd}d^{abcd} = \frac{N_G (N_G-1)(N_G-3)}{12(N_G+2)} \ .
\end{align} 

\newpage
\section{Zerology of the Exceptional Groups}\label{Zerology}
In this appendix is shown the zerology for the exceptional groups with fermions either in the adjoint or fundamental representation.
\begin{figure}[!ht]
	\begin{center}
	\subfloat[$G_2$, Adjoint]{\label{G2Adj} \includegraphics[width=0.45\textwidth]{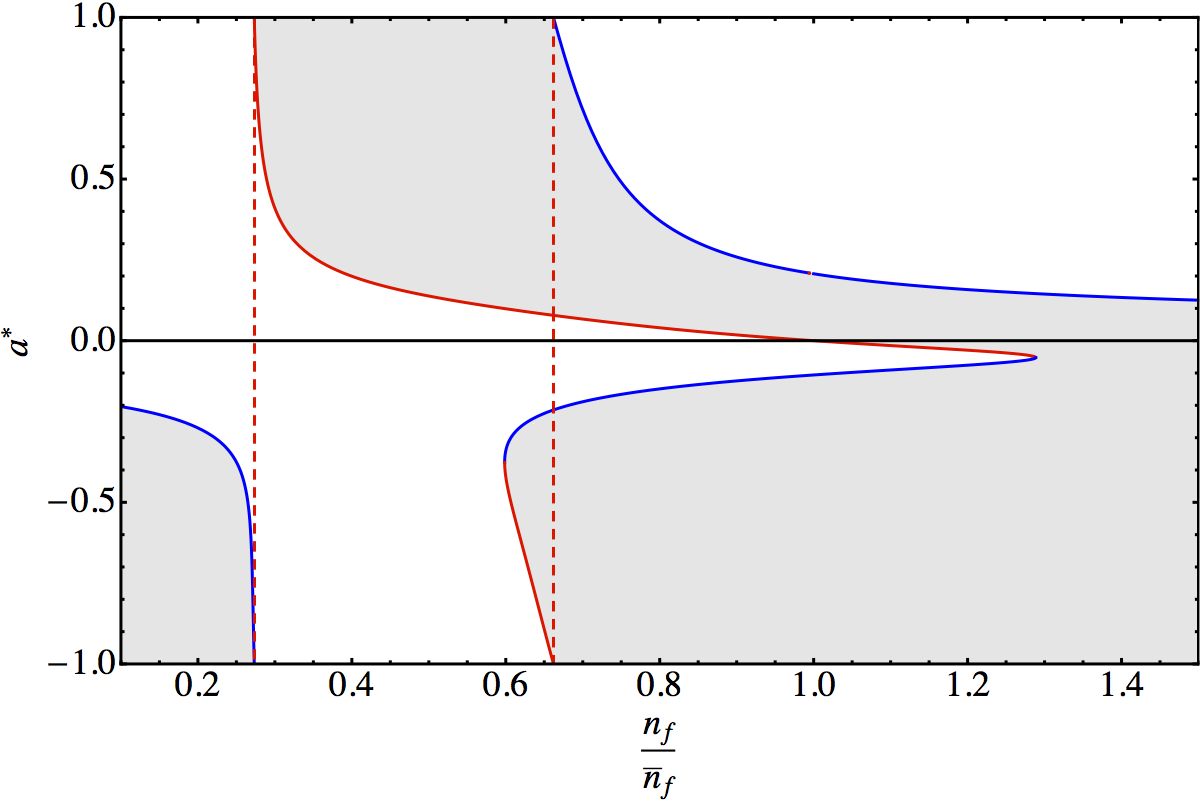}}
	\hfill
	\subfloat[$G_2$, Fundamental]{\label{G2Fund} \includegraphics[width=0.45\textwidth]{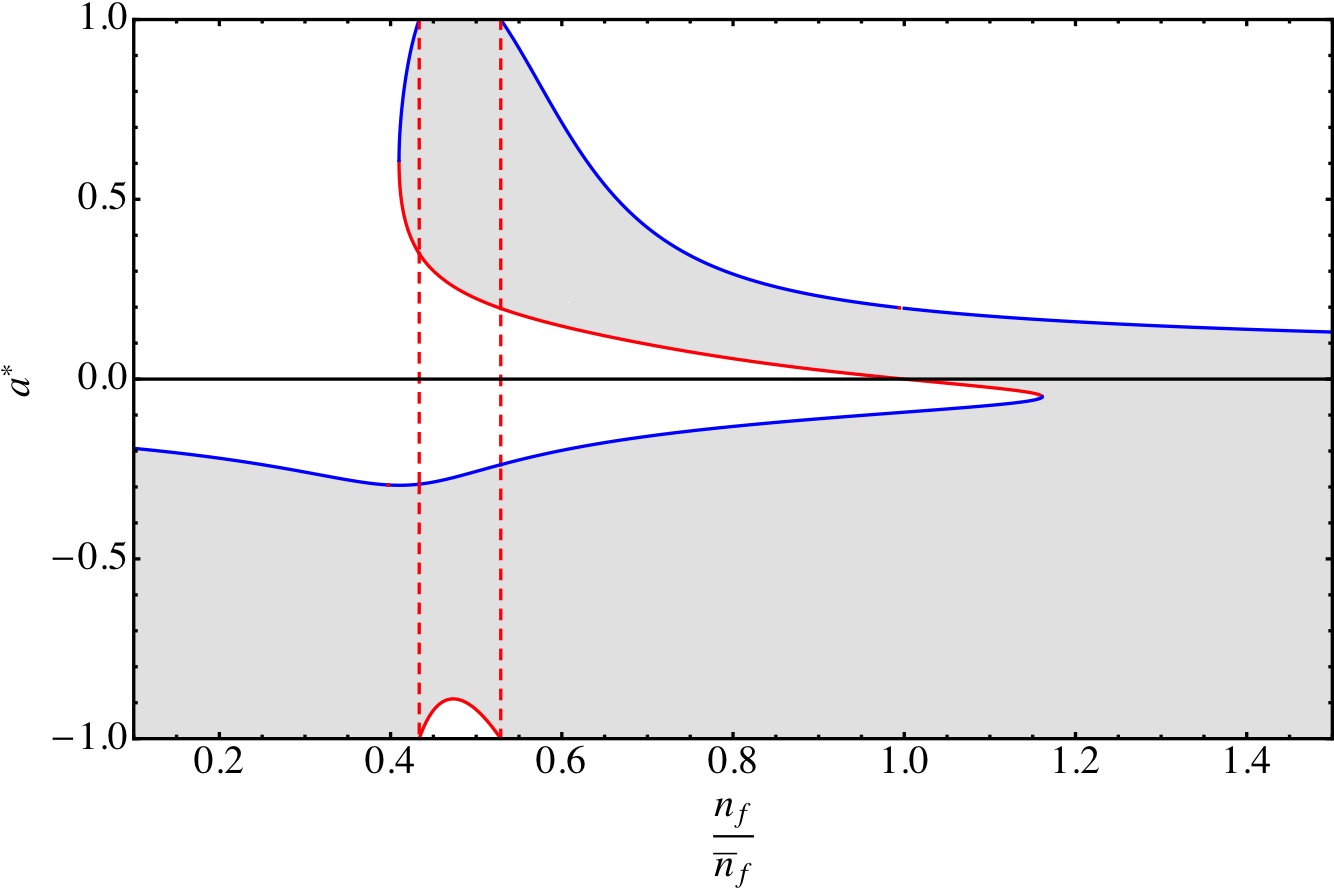}} \\ 
	\subfloat[$F_4$, Adjoint]{\label{F4Adj} \includegraphics[width=0.45\textwidth]{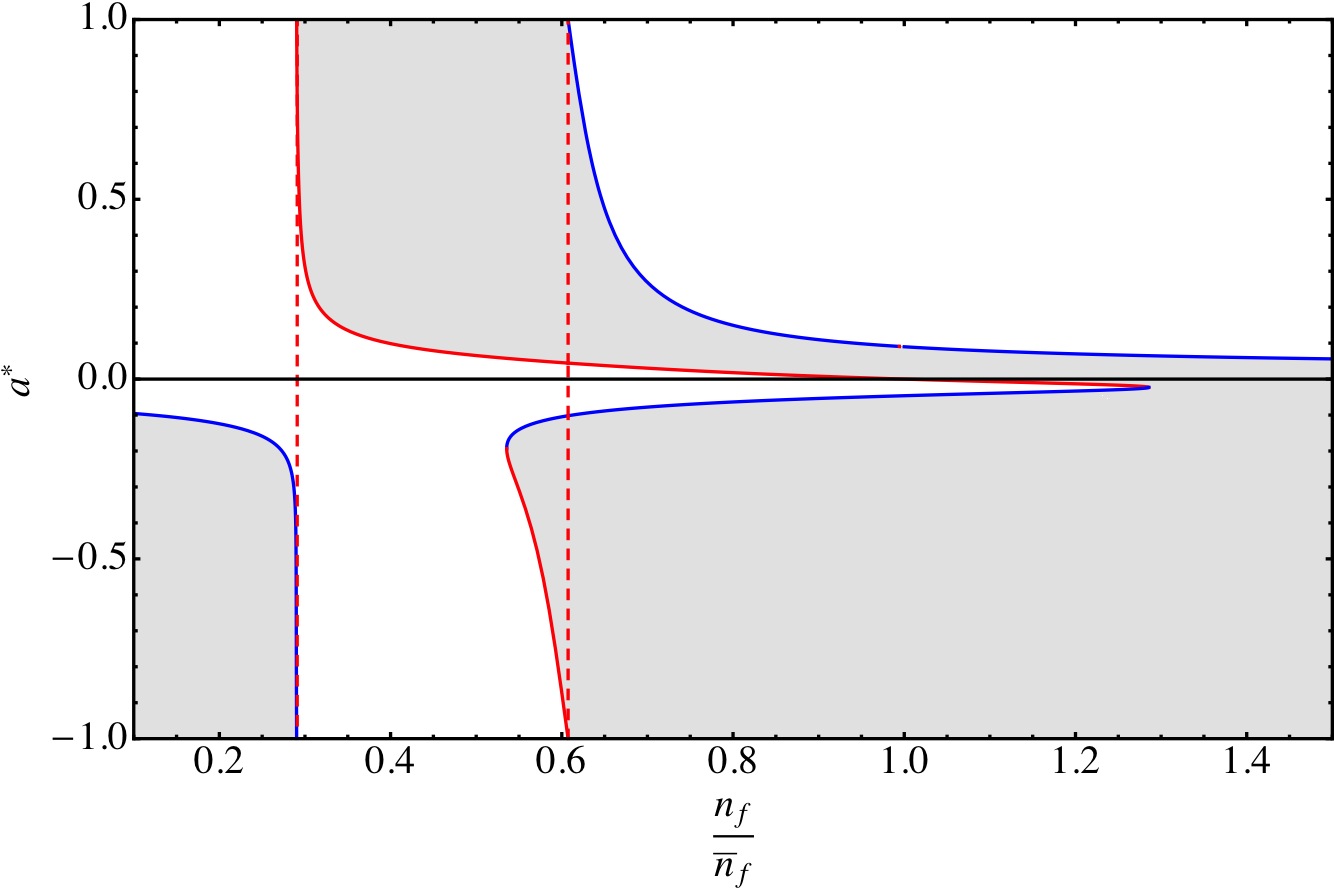}}
	\hfill
	\subfloat[$F_4$, Fundamental]{\label{F4Fund} \includegraphics[width=0.45\textwidth]{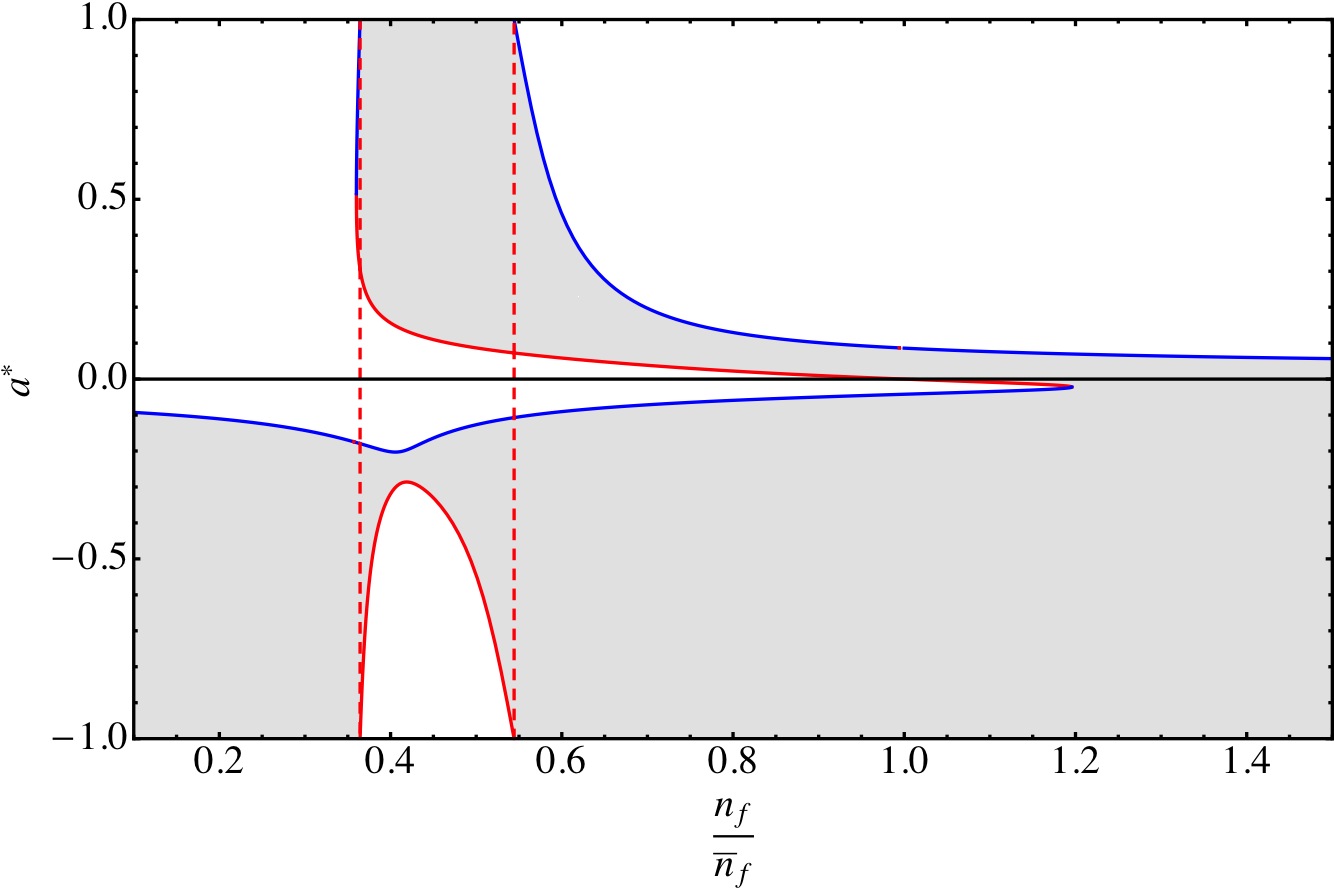}} \\
	\subfloat[$E_6$, Adjoint]{\label{E6Adj} \includegraphics[width=0.45\textwidth]{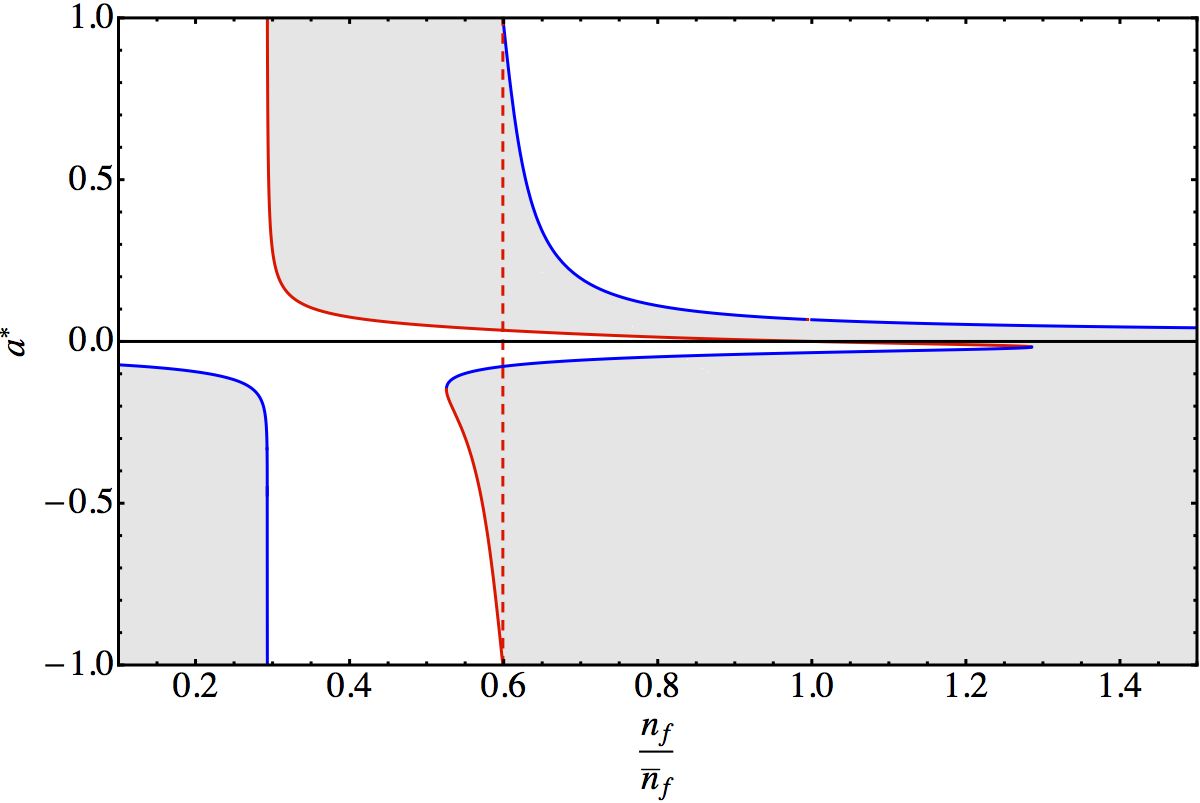}}
	\hfill
	\subfloat[$E_6$, Fundamental]{\label{E6Fund} \includegraphics[width=0.45\textwidth]{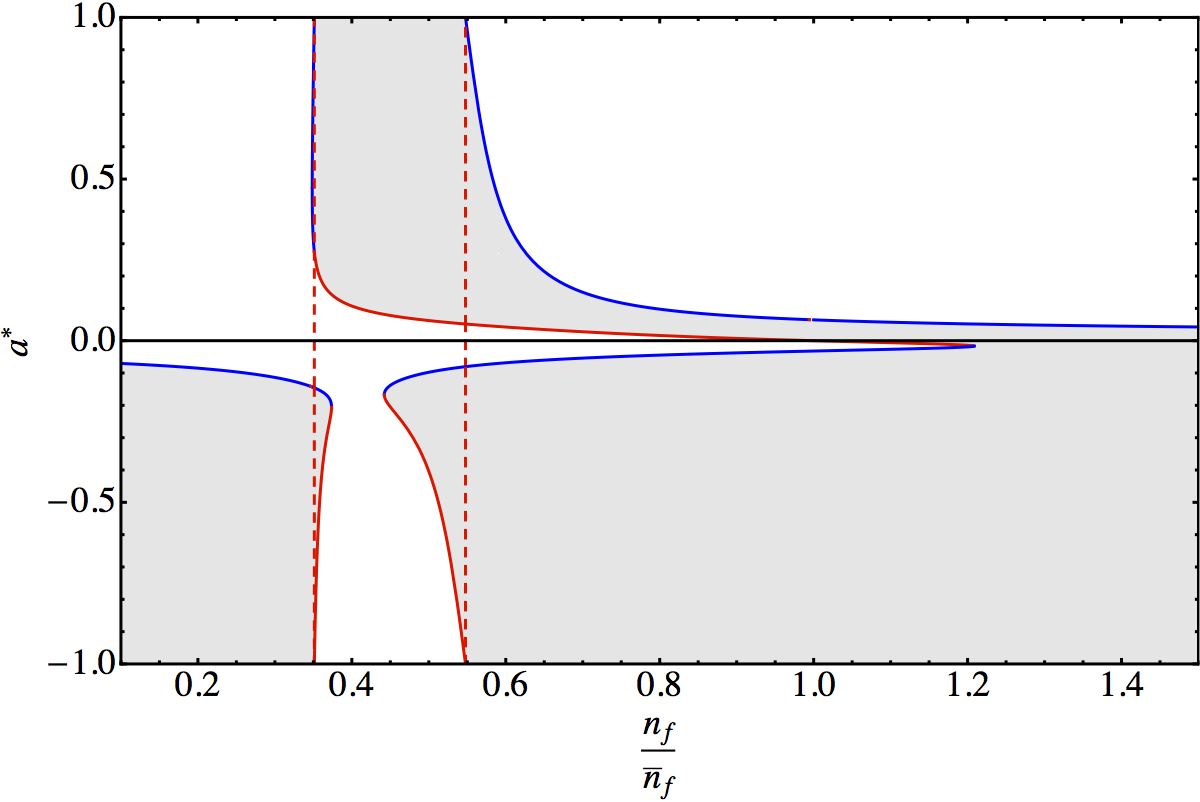}}
	\caption{Zerology to four loops for the given gauge groups and representations.}
	\label{ZerosE6}
	\end{center}
\end{figure}
~
\begin{figure}[h!]
\vspace{-5mm}
	\begin{center}
	\subfloat[$E_7$, Adjoint]{\label{E7Adj} \includegraphics[width=0.45\textwidth]{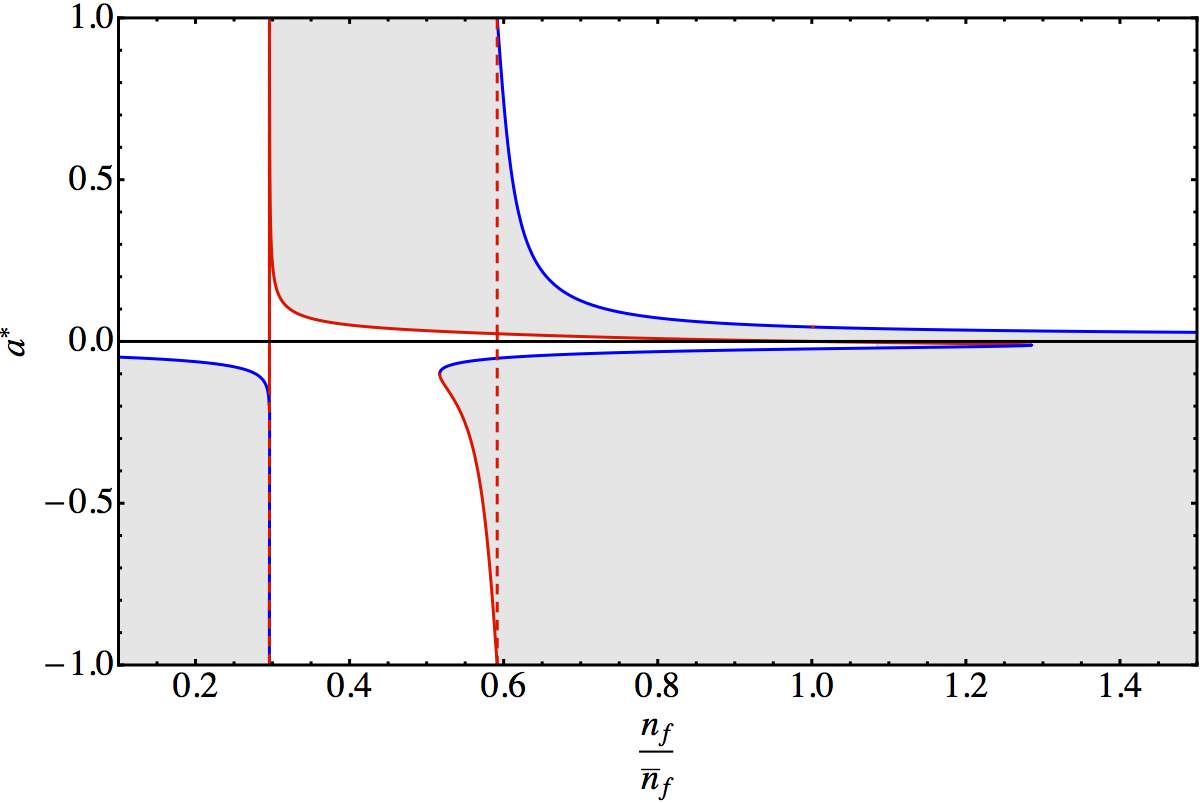}}
	\hfill
	\subfloat[$E_7$, Fundamental]{\label{E7Fund} \includegraphics[width=0.45\textwidth]{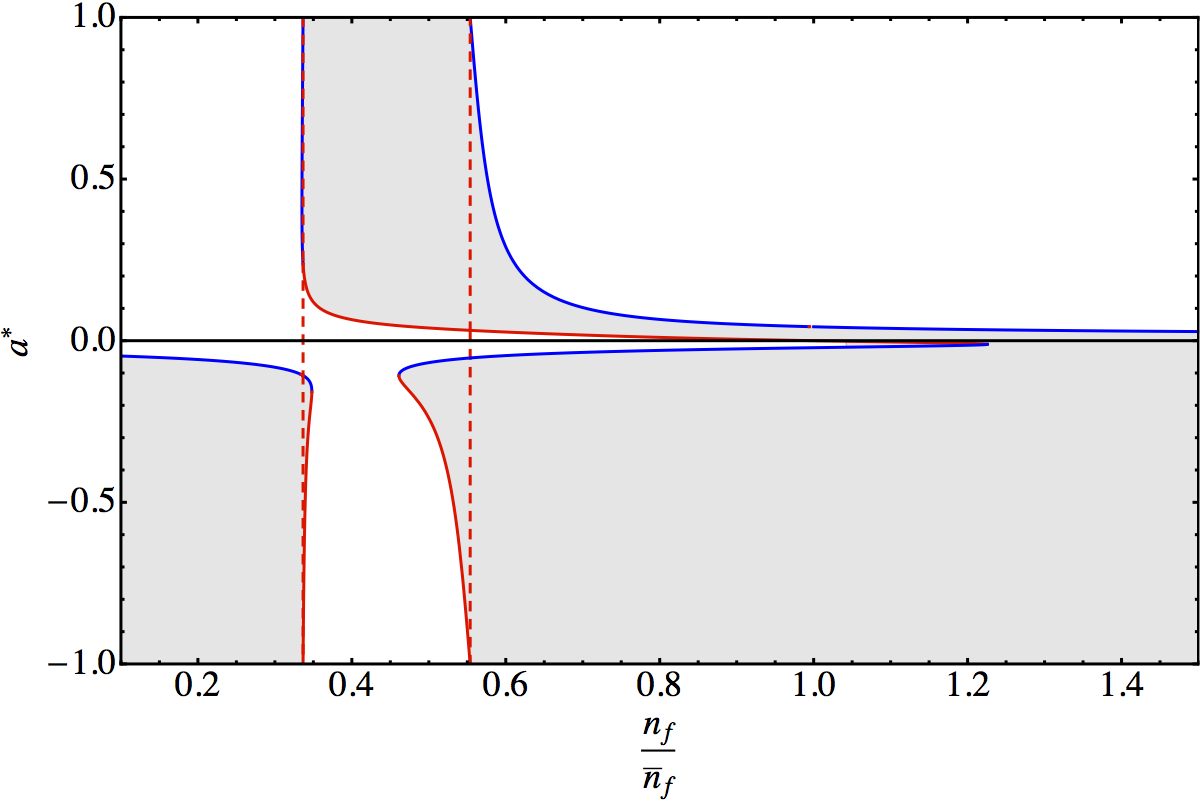}} \\
	 \subfloat[$E_8$, Adjoint/Fundamental]{\label{E8} 
	 \includegraphics[width=0.45\textwidth]{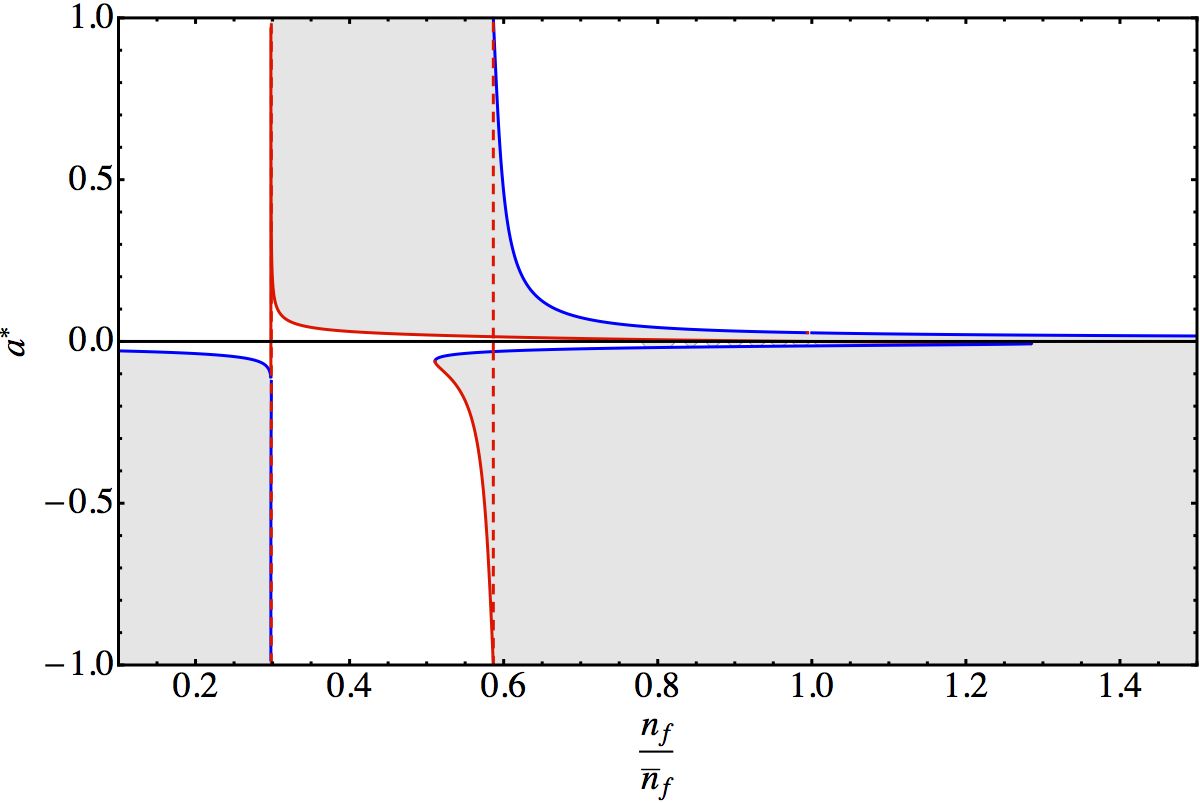}}
	\end{center}
	\vspace{-8mm}
	\caption{Zerology to four loops for the given gauge groups and representations.}
	\vspace{-2mm}
\end{figure}

\section{Anomalous Dimension for the Exceptional Groups and Spinorial Representations}\label{Gamma}
\begin{figure}[!h]
\vspace{-2mm}
	\begin{center}
	\subfloat[$G_2$, Adjoint]{\label{GammaG2Adj} \includegraphics[width=0.45\textwidth]{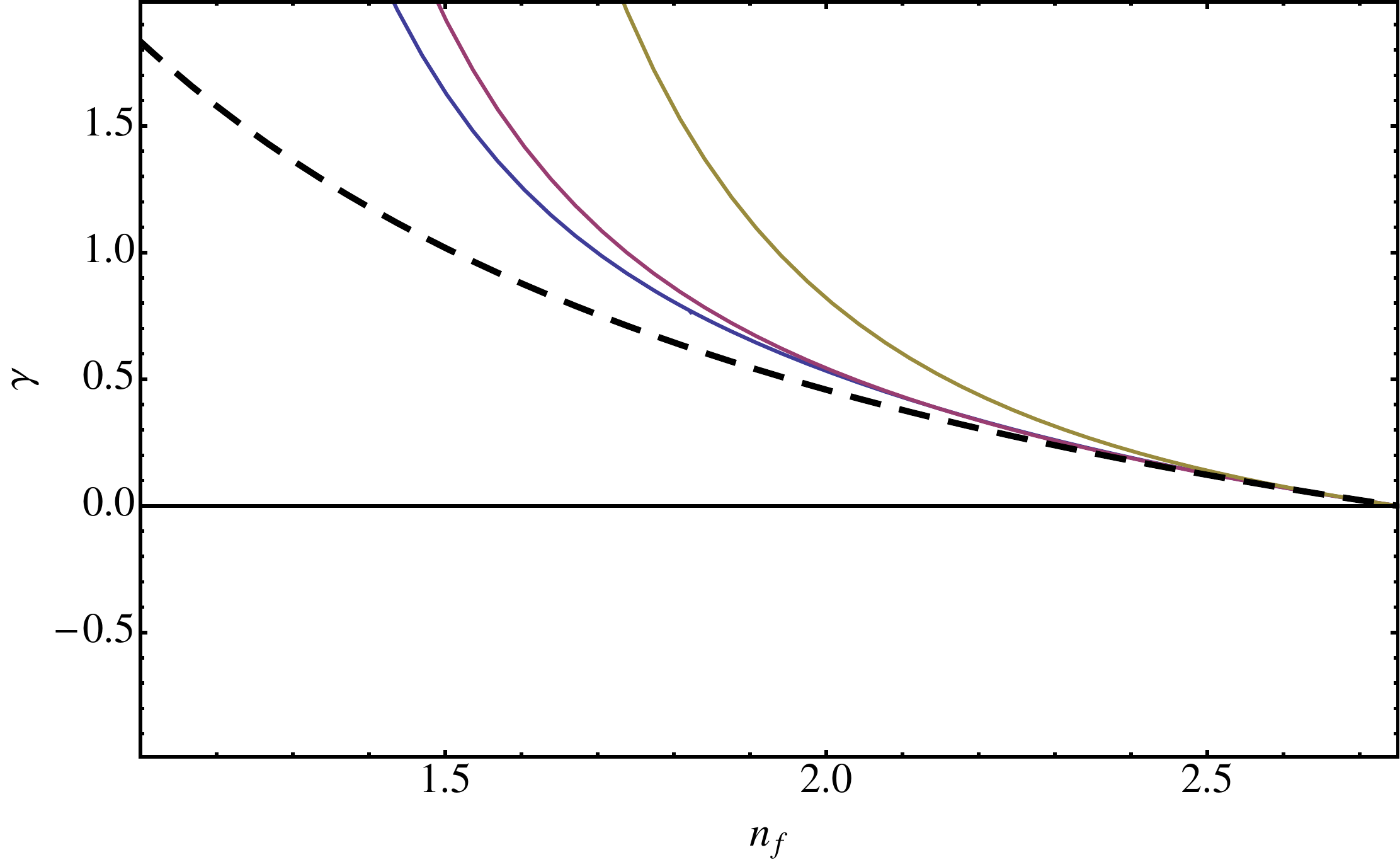}}
	\hfill
	\subfloat[$G_2$, Fundamental]{\label{GammaG2Fund} \includegraphics[width=0.45\textwidth]{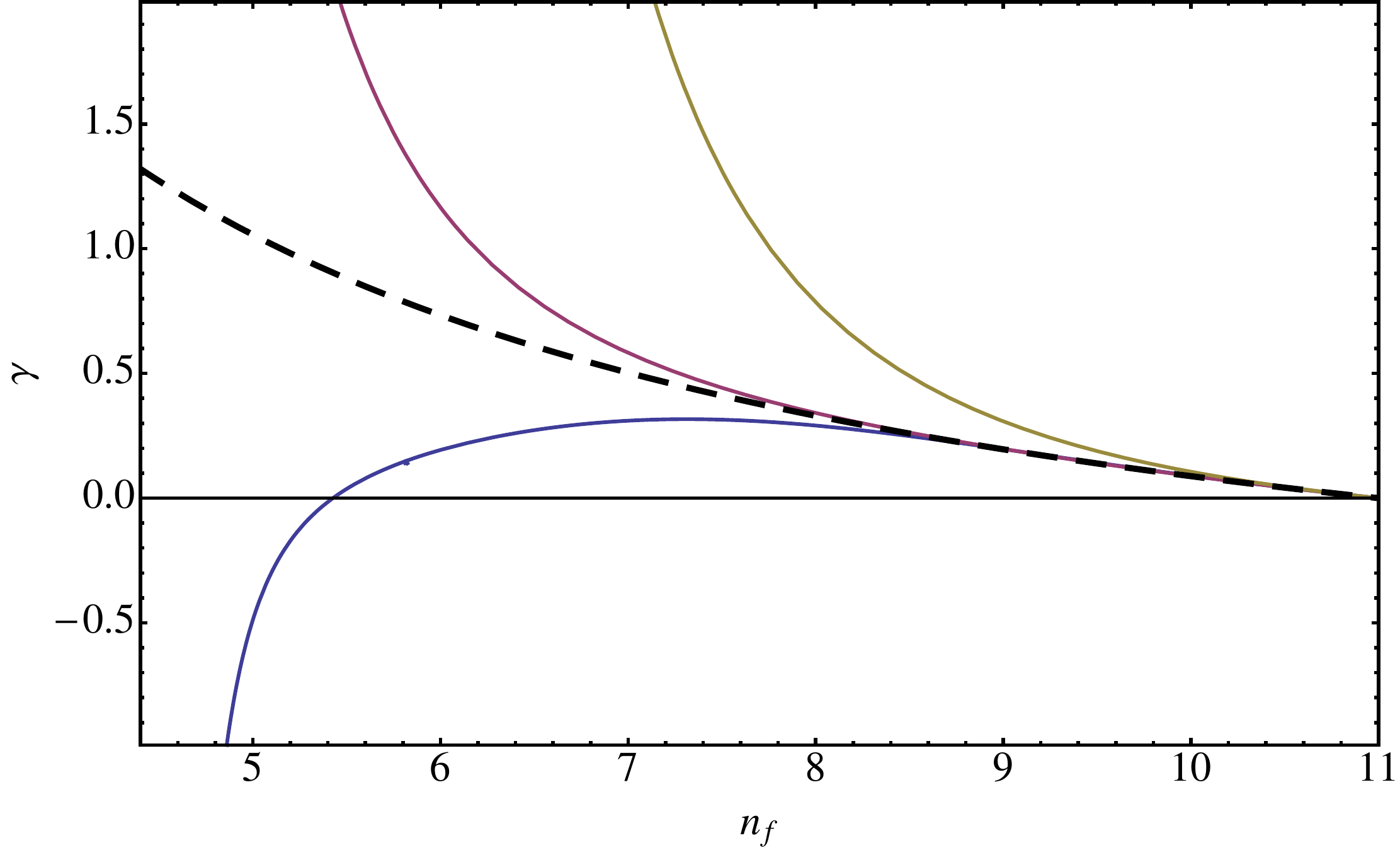}}
	\vspace{-3mm}
\caption{Anomalous dimension at the perturbative fixed point as a function of number of flavors $n_f$. The curve corresponding to i) two loops is yellow, ii) three loops is red, iii) four loops is blue iv) all-orders is dashed. }\label{Gamma1}
\vspace{-5mm}
\end{center}
\end{figure}

\begin{figure}[h!]
\vspace{-5mm}
 \begin{center}
	\subfloat[$F_4$, Adjoint]{\label{GammaF4Adj} \includegraphics[width=0.45\textwidth]{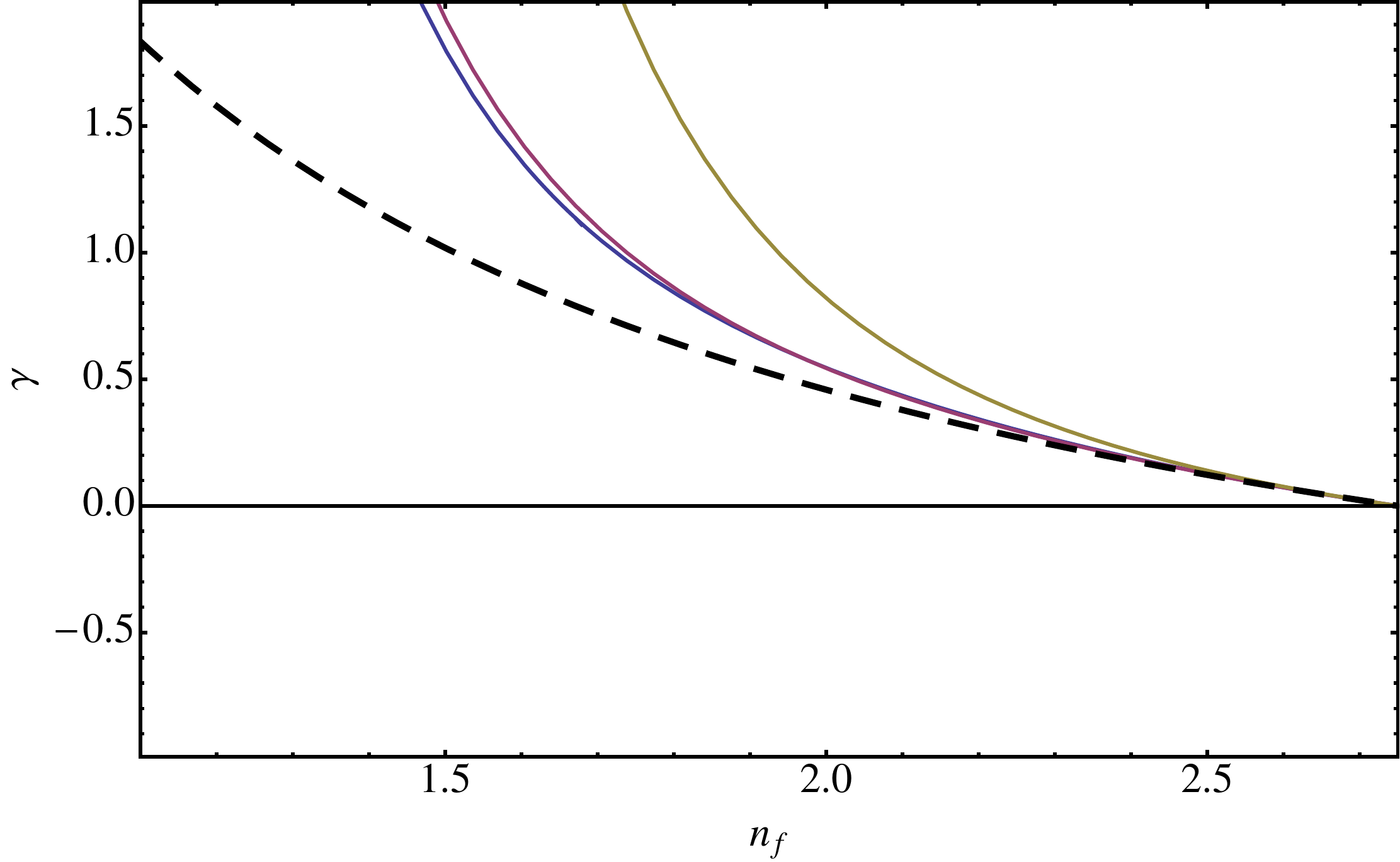}}
	\hfill
	\subfloat[$F_4$, Fundamental]{\label{GammaF4Fund} \includegraphics[width=0.45\textwidth]{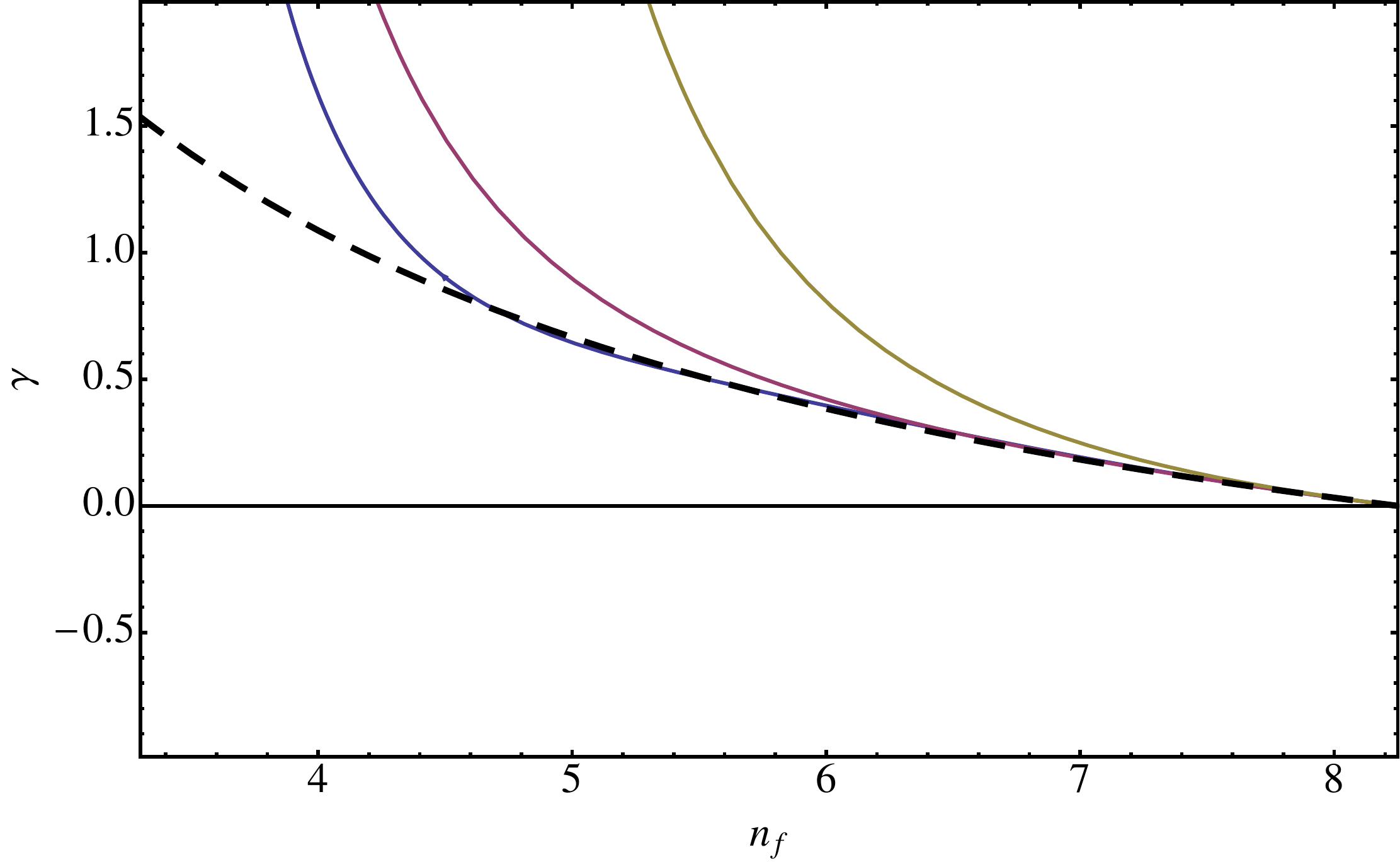}}\\
	\subfloat[$E_6$, Adjoint]{\label{GammaE6Adj} \includegraphics[width=0.45\textwidth]{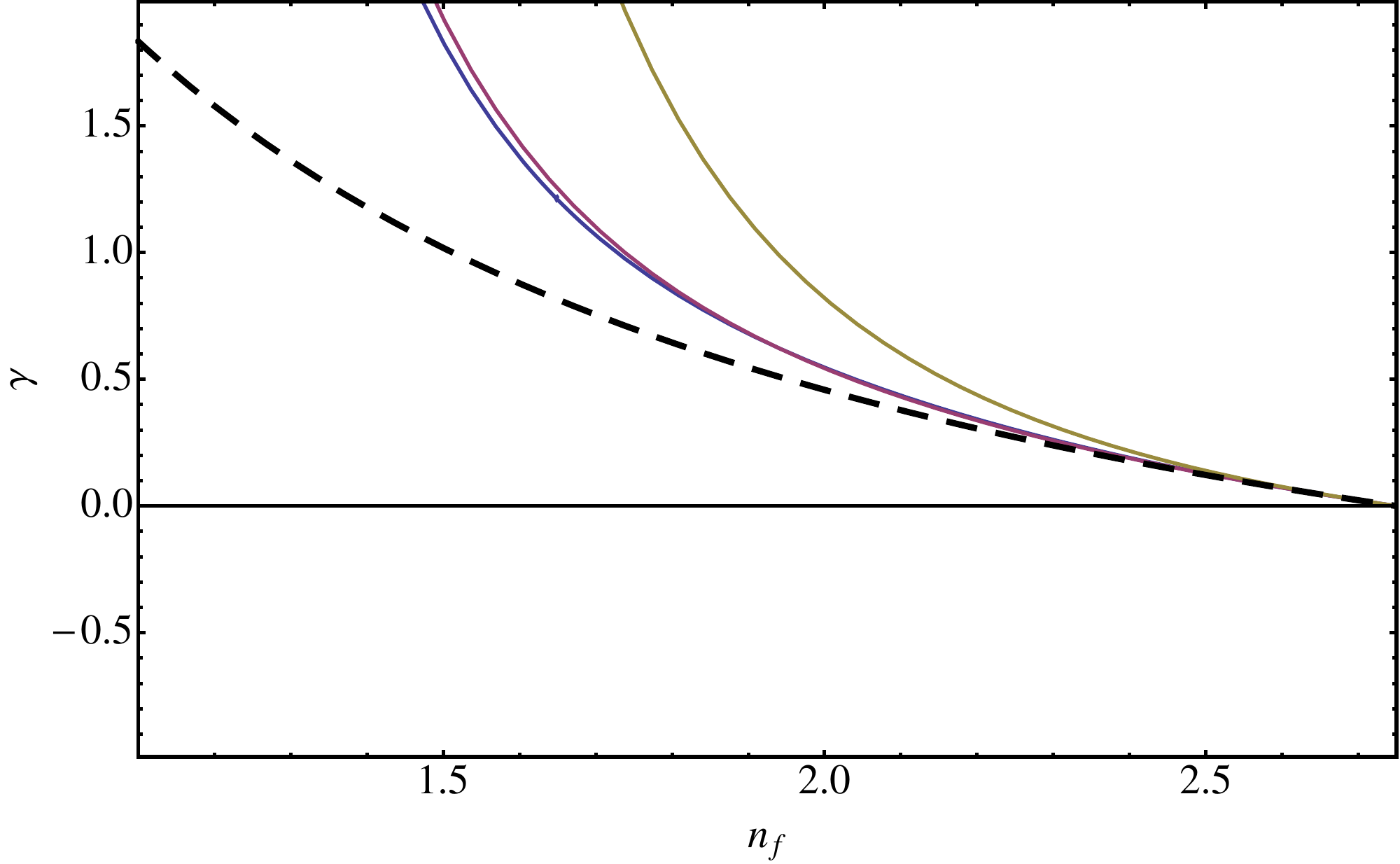}}
	\hfill
	\subfloat[$E_6$, Fundamental]{\label{GammaE6Fund} \includegraphics[width=0.45\textwidth]{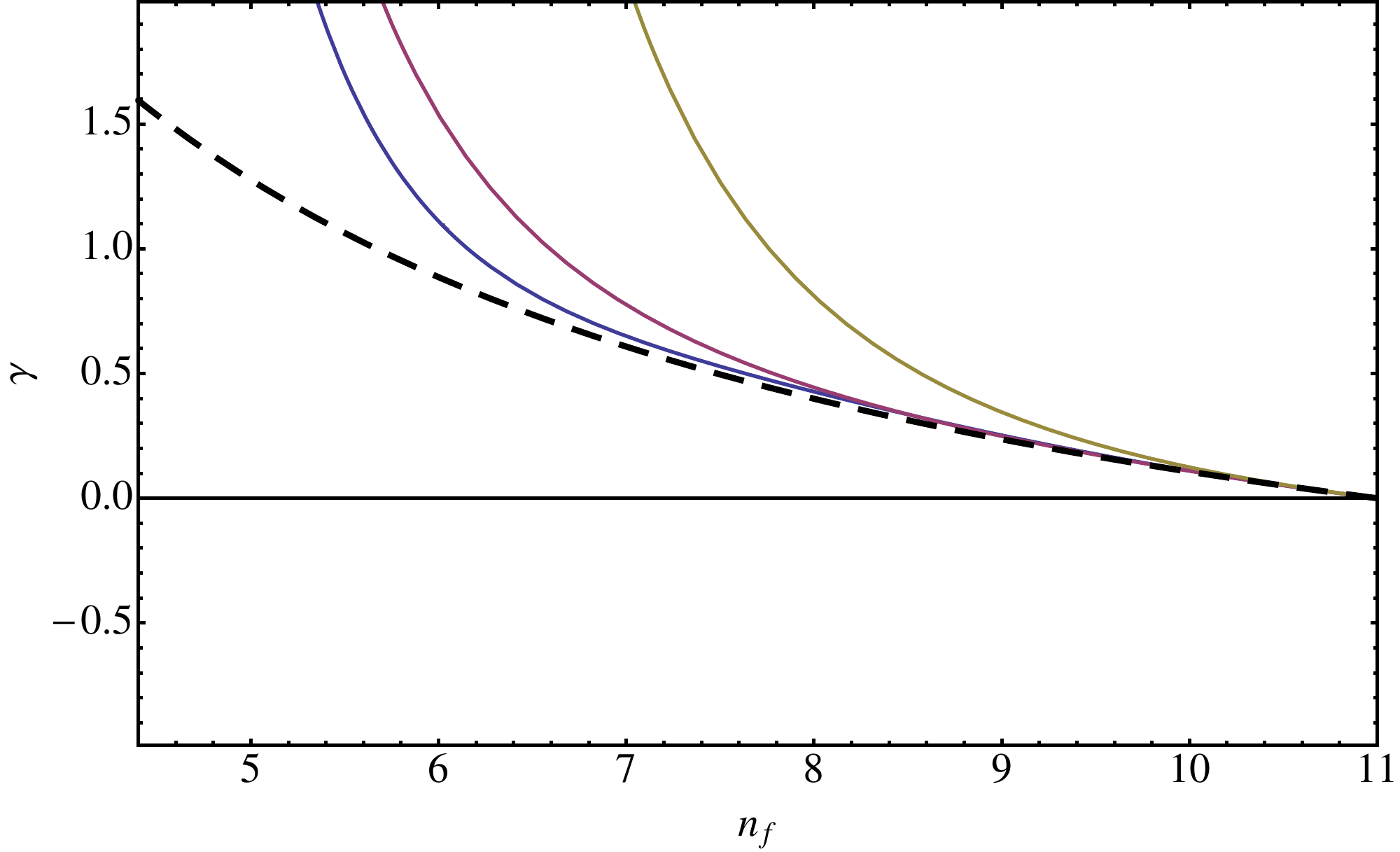}} \\
	\subfloat[$E_7$, Adjoint]{\label{GammaE7Adj} \includegraphics[width=0.45\textwidth]{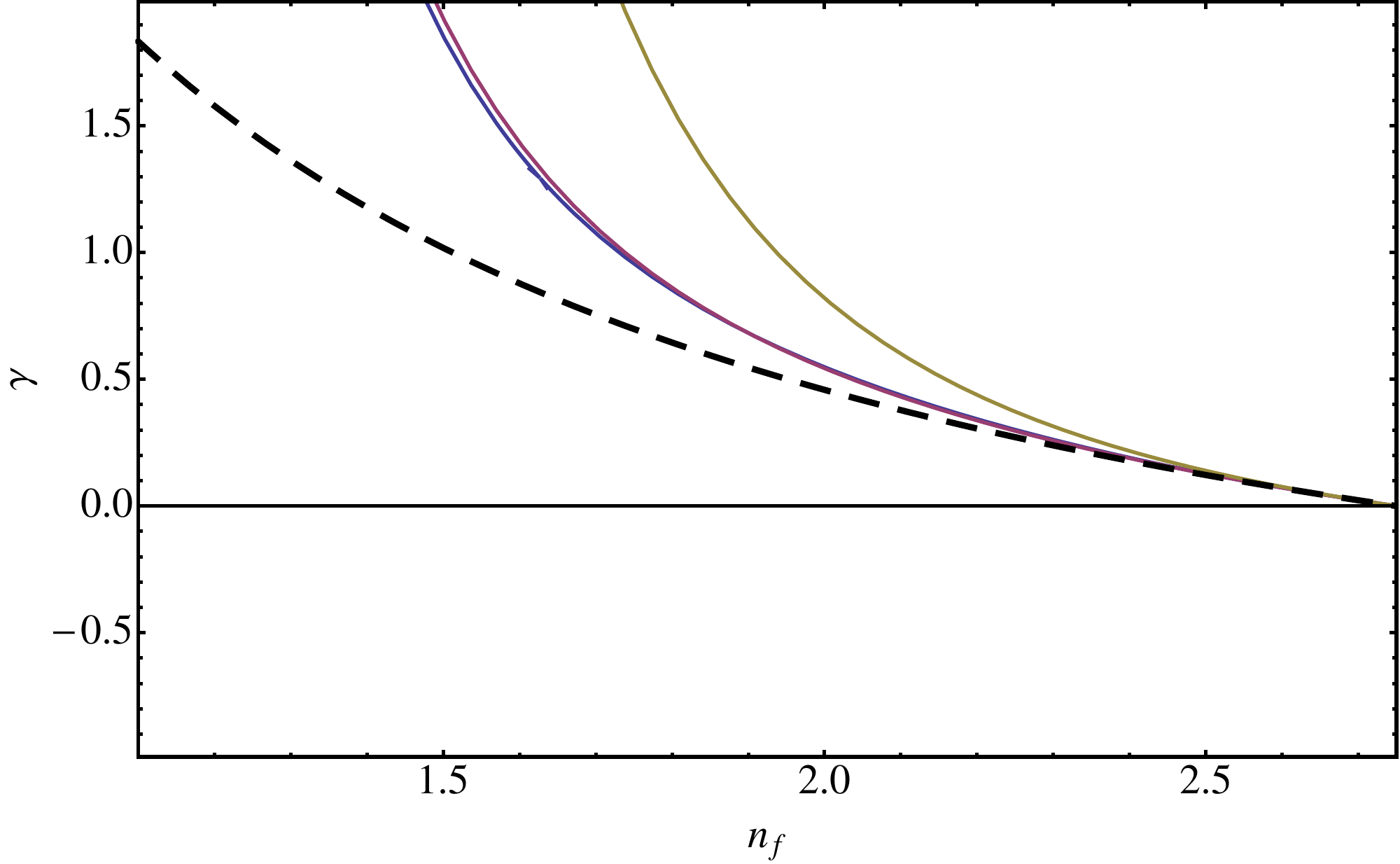}}
	\hfill
	\subfloat[$E_7$, Fundamental]{\label{GammaE7Fund} \includegraphics[width=0.45\textwidth]{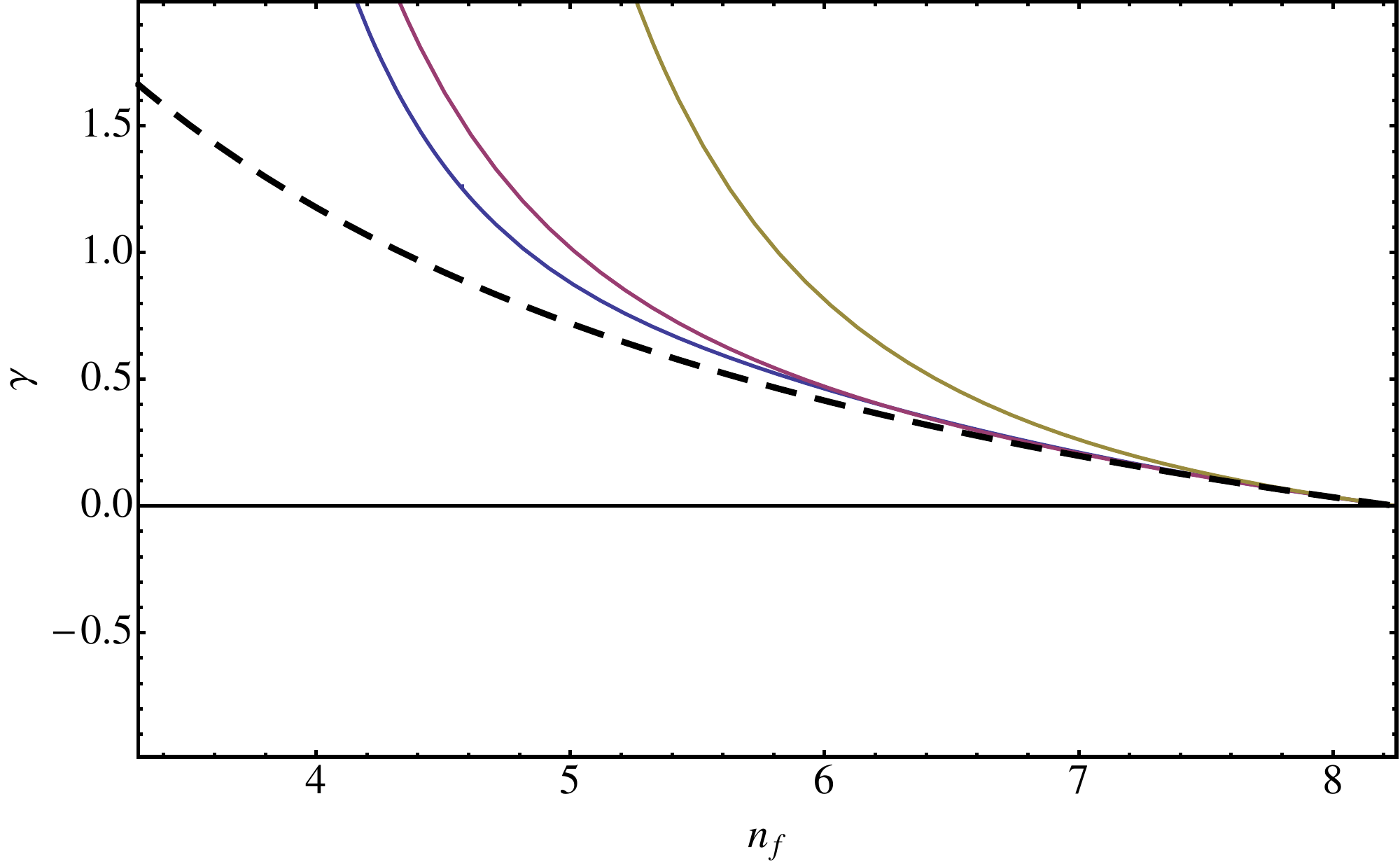}} \\
	 \subfloat[$E_8$, Adjoint/Fundamental]{\label{GammaE8Fund}
	 \includegraphics[width=0.45\textwidth]{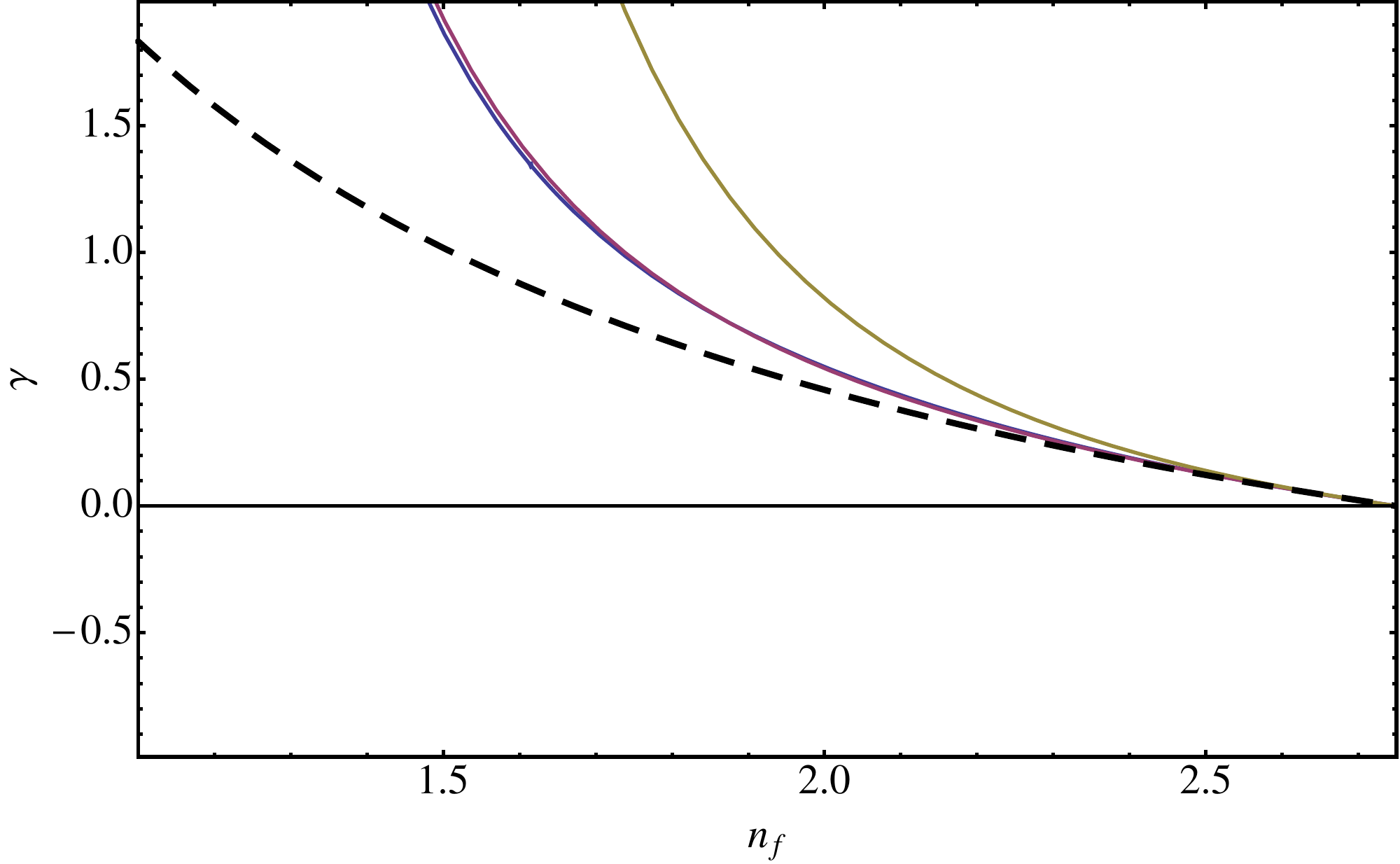}}
	 \vspace{-3mm}
	\caption{Anomalous dimension at the perturbative fixed point as a function of number of flavors $n_f$. Legend as in Figure \ref{Gamma1}.}
	\label{Gamma2}
	\end{center}
\end{figure}

\vspace{-4mm}
We provide a plot of the anomalous dimension at the IR fixed point stemming from  the two, three and four loop beta function as well as the all-order beta function.  For the exceptional groups these are reported in Fig \ref{Gamma1} an Fig.~\ref{Gamma2}. 

\pagebreak

\FloatBarrier

Similarly we provide below a plot for the anomalous dimension for some representatives of the spinorial representations as clearly labeled in the associated figures.
\begin{figure}[h]
	\begin{center}
	\subfloat[$SO(5)$]{\label{GammaSO5} 
	\includegraphics[width=0.49\textwidth]{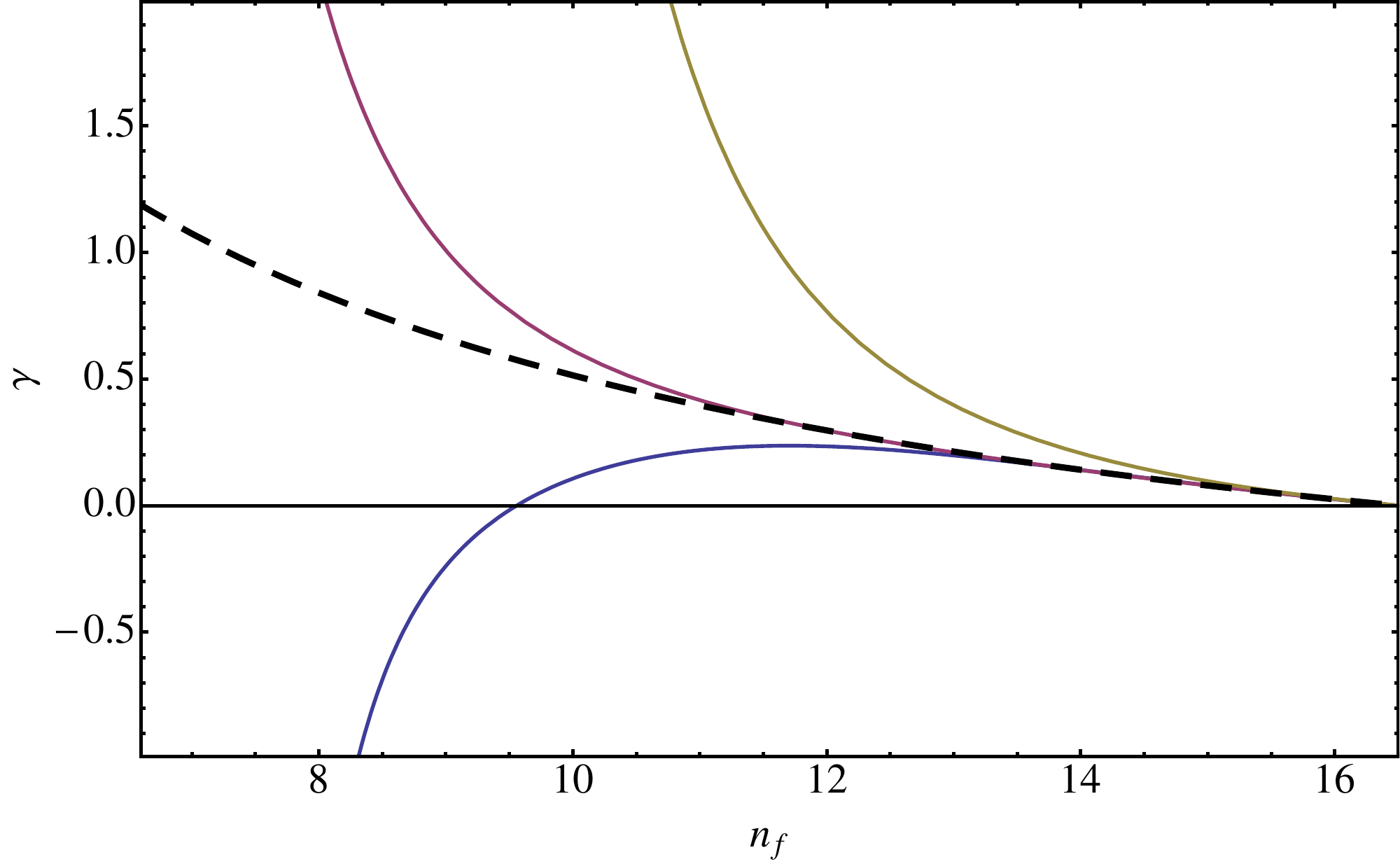}
	}
	%
	\subfloat[$SO(8)$]{\label{GammaSO8} 
	\includegraphics[width=0.49\textwidth]{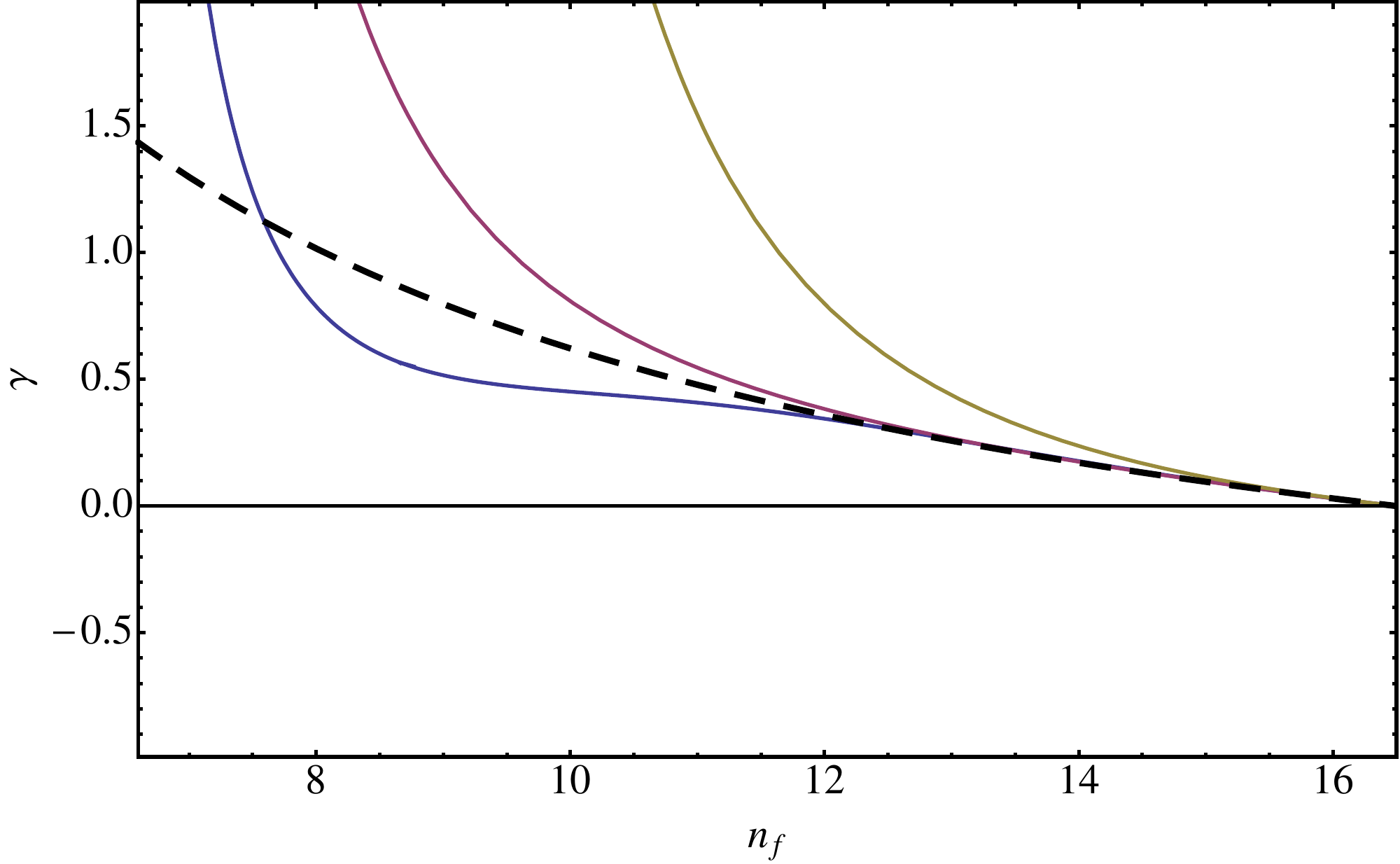}
	}\\
	\subfloat[$SO(10)$]{\label{GammaSO10} 
	\includegraphics[width=0.49\textwidth]{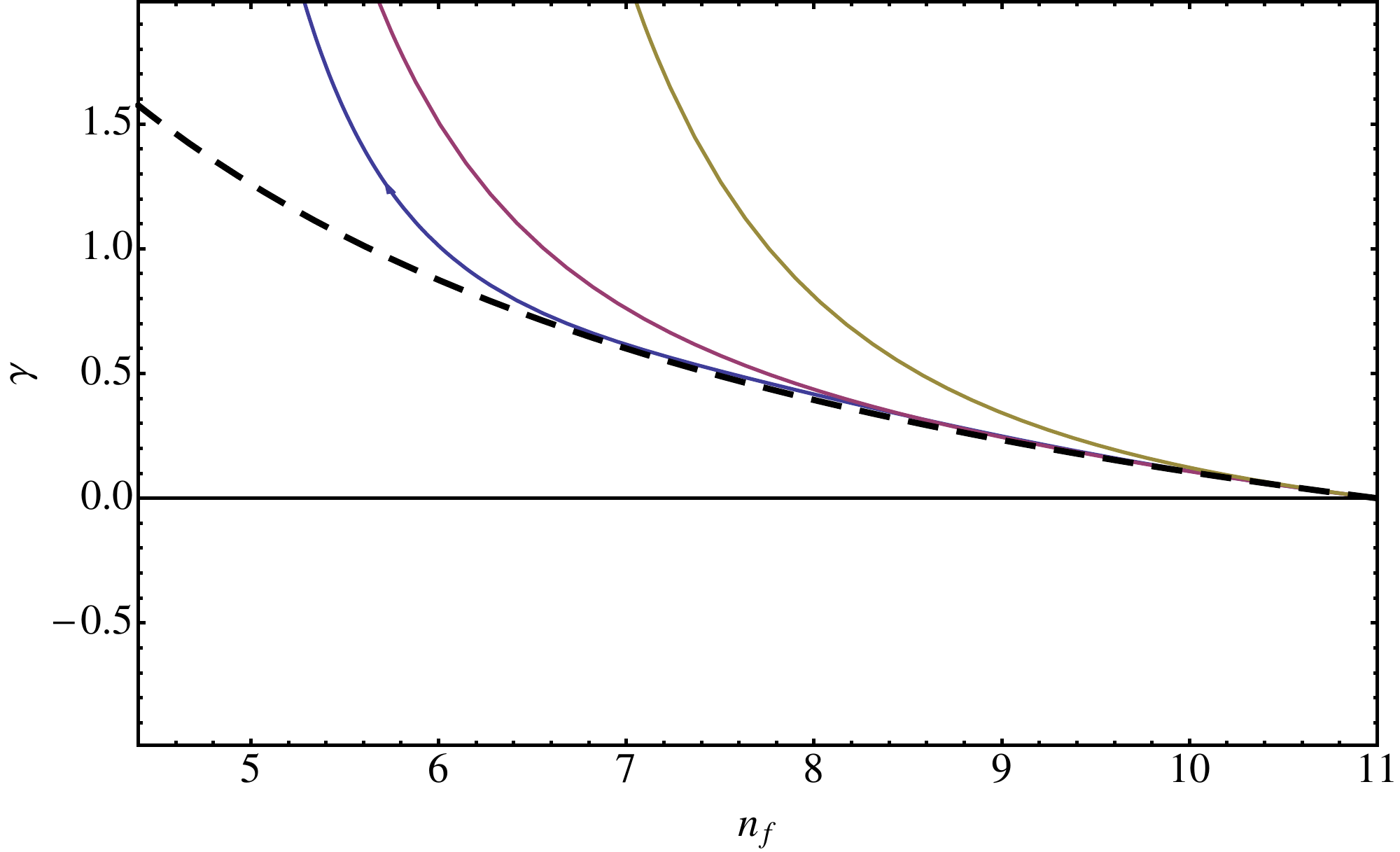}
	}
	\subfloat[$SO(12)$]{\label{GammaSO12} 
	\includegraphics[width=0.49\textwidth]{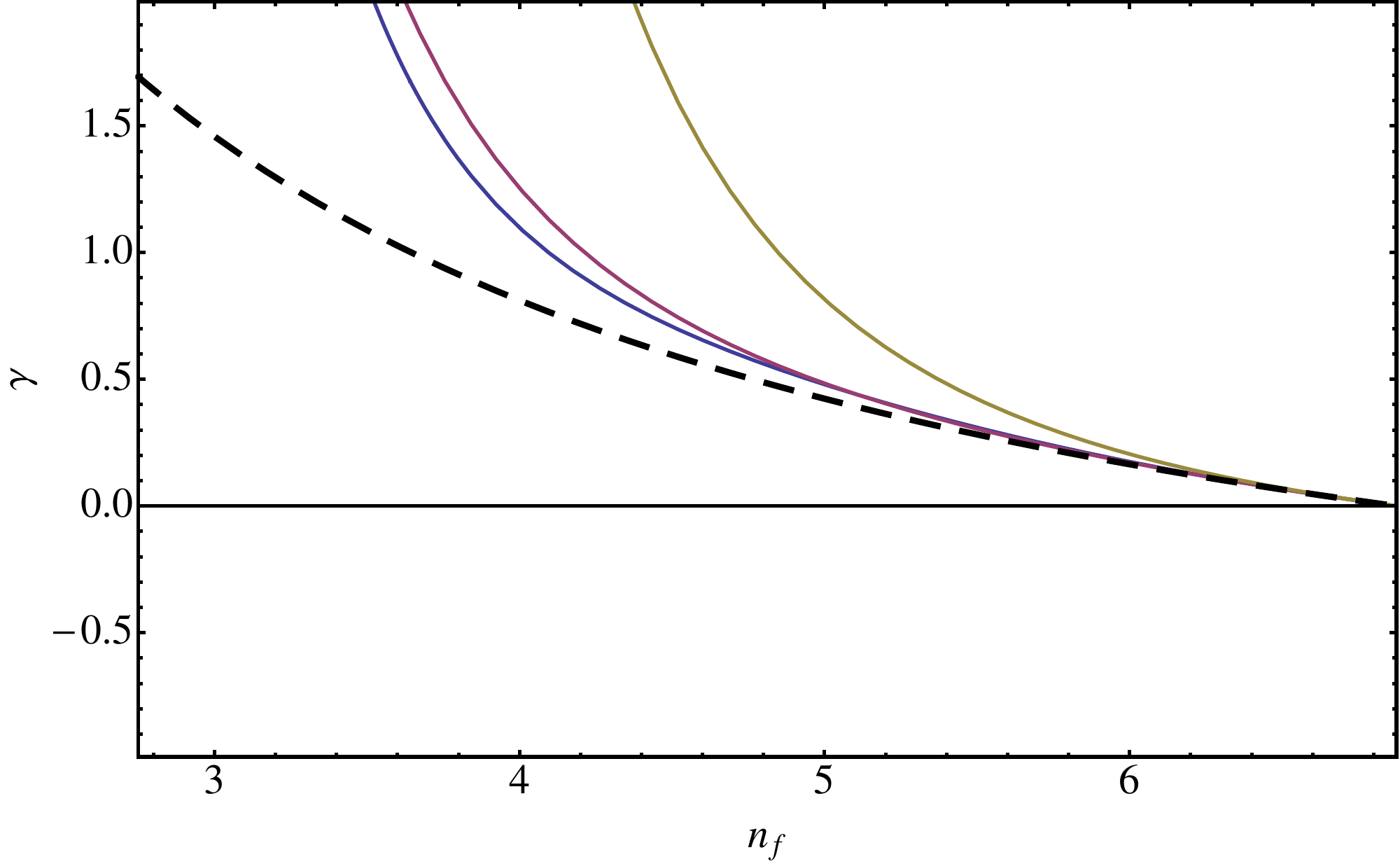}
	}
	\caption{Anomalous dimension at the perturbative fixed point as a function of number of flavors $n_f$ for some $SO(N)$ gauge group, illustrating the trend as $N$ increases. The curve corresponding to i) two-loops is yellow, ii) three-loops is red, iii) four-loops is blue iv) all-orders is dashed.}
	\label{GammaSON}
	\end{center}
\end{figure}

We finally summarize the numerical values of 
the fixed point and anomalous dimension for various gauge theories with different number of flavors to two, three and four loops as well as the all orders result in Table \ref{TBLExceptional}-\ref{TBLSpin}.

\begin{table}[h]
\begin{tabular}{|c|ccc|cccc|}
\hline\multicolumn{8}{|c|}{$G_2$ Fundamental} \\ \hline
$n_f$ &  $a_ 2^*$ & $a_ 3^*$ & $a_ 4^*$ & $\gamma _ 2^*$ & $\gamma_ 3^*$ & $\gamma _ 4^*$ & $\gamma _ {\text{AO}}^*$\\ 
 5. & -1. & 0.102 & 0.098  & 192. & 3.821 & -0.488  & 1.056 \\
 6. & 0.25 & 0.058  & 0.059 & 14.917 & 1.165  & 0.193  & 0.733 \\
 7. & 0.087 & 0.038  & 0.041 & 2.384 & 0.582  & 0.31  & 0.503 \\
 8. & 0.042 & 0.025  & 0.027 & 0.785 & 0.341  & 0.291  & 0.33 \\
 9. & 0.02 & 0.015  & 0.016  & 0.308 & 0.197  & 0.196  & 0.196 \\
 10. & 0.008 & 0.007  & 0.007 & 0.106 & 0.089  & 0.09 & 0.088 \\
 11. & 0 & 0  & 0 & 0 & 0  & 0 & 0 \\
\hline
\end{tabular}
\hfill
\begin{tabular}{|c|ccc|cccc|}
\hline\multicolumn{8}{|c|}{$F_4$ Fundamental} \\ \hline
$n_f$ &  $a_ 2^*$ & $a_ 3^*$ & $a_ 4^*$ & $\gamma _ 2^*$ & $\gamma_ 3^*$ & $\gamma _ 4^*$ & $\gamma _ {\text{AO}}^*$\\ 
 4. & 0.189 & 0.031  & 0.029 & 55.823 & 2.763  & 1.623 & 1.087 \\
 5. & 0.038 & 0.017 & 0.018  & 3.18 & 0.897 & 0.645  & 0.665 \\
 6. & 0.015 & 0.01 & 0.011  & 0.806 & 0.42 & 0.396  & 0.384 \\
 7. & 0.006 & 0.005 & 0.005  & 0.248 & 0.189 & 0.192  & 0.183 \\
 8. & 0.001 & 0.001 & 0.001  & 0.034 & 0.032 & 0.032  & 0.032 \\
\hline
\end{tabular} \\
\vspace{5mm}
\begin{tabular}{|c|ccc|cccc|}
\hline\multicolumn{8}{|c|}{$E_6$ Fundamental} \\ \hline
$n_f$ &  $a_ 2^*$ & $a_ 3^*$ & $a_ 4^*$ & $\gamma _ 2^*$ & $\gamma_ 3^*$ & $\gamma _ 4^*$ & $\gamma _ {\text{AO}}^*$\\ 
 5. & 0.273 & 0.027 & 0.024  & 216.595 & 4.393 & 3.153  & 1.276 \\
 6. & 0.046 & 0.016 & 0.016  & 7.869 & 1.536 & 1.11  & 0.886 \\
 7. & 0.021 & 0.011  & 0.011  & 2.082 & 0.775 & 0.65  & 0.608 \\
 8. & 0.011 & 0.007  & 0.008  & 0.81 & 0.444 & 0.427 & 0.399 \\
 9. & 0.005 & 0.004  & 0.004 & 0.345 & 0.249  & 0.252  & 0.236 \\
 10. & 0.002 & 0.002  & 0.002  & 0.124 & 0.109  & 0.11 & 0.106 \\
 11. & 0 & 0  & 0 & 0 & 0  & 0 & 0 \\
\hline
\end{tabular}
\hfill
\begin{tabular}{|c|ccc|cccc|}
\hline\multicolumn{8}{|c|}{$E_7$ Fundamental} \\ \hline
$n_f$ &  $a_ 2^*$ & $a_ 3^*$ & $a_ 4^*$ & $\gamma _ 2^*$ & $\gamma_ 3^*$ & $\gamma _ 4^*$ & $\gamma _ {\text{AO}}^*$\\ 
 4. & 0.059 & 0.014 & 0.013  & 28.122 & 3.049 & 2.483 & 1.178 \\
 5. & 0.016 & 0.008 & 0.008  & 2.882 & 1.019 & 0.881  & 0.721 \\
 6. & 0.007 & 0.005 & 0.005  & 0.814 & 0.471 & 0.463  & 0.416 \\
 7. & 0.003 & 0.002 & 0.002  & 0.262 & 0.208 & 0.212  & 0.198 \\
 8. & 0.0004 & 0.0004  & 0.0004 & 0.0362 & 0.0350  & 0.0351 & 0.0346 \\
\hline
\end{tabular}
\caption{Comparison between different determinations of the anomalous dimension of the mass for exceptional gauge theories with $n_f$ fermions in the fundamental representation. The anomalous dimensions $\gamma^*_2$, $\gamma^*_3$ and $\gamma^*_4$ are the perturbative result at 2, 3 and 4-loop respectively, while $\gamma^*_{AO}$ corresponds to the all orders result. We also report the corresponding value of the zero of the $\beta$ function ($a^*=\alpha^*/4\pi$ ), at 2, 3 and 4 loops respectively and indicated with $a^*_2$, $a^*_3$ and $a^*_4$ .\label{TBLExceptional}}
\end{table}

\begin{table}[h]
\begin{tabular}{|c|ccc|cccc|}
\hline\multicolumn{8}{|c|}{$G_2$ Adjoint} \\ \hline
$n_f$ &  $a_ 2^*$ & $a_ 3^*$ & $a_ 4^*$ & $\gamma _ 2^*$ & $\gamma_ 3^*$ & $\gamma _ 4^*$ & $\gamma _ {\text{AO}}^*$\\ 
 1.5 & 0.089 & 0.041  & 0.038 & 5.374 & 1.924  & 1.633 & 1.019\\
 1.75 & 0.045 & 0.028  & 0.027 & 1.873 & 0.972 & 0.895 & 0.698 \\
 2. & 0.025 & 0.018  & 0.019 & 0.82 & 0.543& 0.532  & 0.458 \\
 2.25 & 0.013 & 0.011 & 0.011  & 0.372 & 0.296 & 0.298  & 0.272 \\
 2.5 & 0.005 & 0.005 & 0.005  & 0.139 & 0.127 & 0.128  & 0.122 \\
 2.75 & 0 & 0  & 0 & 0 & 0  & 0 & 0 \\
\hline
\end{tabular}
\hfill
\begin{tabular}{|c|ccc|cccc|}
\hline\multicolumn{8}{|c|}{$F_4$ Adjoint} \\ \hline
$n_f$ &  $a_ 2^*$ & $a_ 3^*$ & $a_ 4^*$ & $\gamma _ 2^*$ & $\gamma_ 3^*$ & $\gamma _ 4^*$ & $\gamma _ {\text{AO}}^*$\\ 
 1.5 & 0.04 & 0.018  & 0.018 & 5.374 & 1.924 & 1.802 & 1.019 \\
 1.75 & 0.02 & 0.012  & 0.013  & 1.873 & 0.972  & 0.942  & 0.698 \\
 2. & 0.011 & 0.008  & 0.008  & 0.82 & 0.543  & 0.545 & 0.458 \\
 2.25 & 0.006 & 0.005  & 0.005  & 0.372 & 0.296 & 0.3 & 0.272 \\
 2.5 & 0.002 & 0.002  & 0.002  & 0.139 & 0.127  & 0.128  & 0.122 \\
 2.75 & 0 & 0  & 0 & 0 & 0  & 0 & 0 \\
\hline
\end{tabular} \\
\vspace{5mm}
\begin{tabular}{|c|ccc|cccc|}
\hline\multicolumn{8}{|c|}{$E_6$ Adjoint} \\ \hline
$n_f$ &  $a_ 2^*$ & $a_ 3^*$ & $a_ 4^*$ & $\gamma _ 2^*$ & $\gamma_ 3^*$ & $\gamma _ 4^*$ & $\gamma _ {\text{AO}}^*$\\ 
 1.5 & 0.03 & 0.014  & 0.013 & 5.374 & 1.924  & 1.83 & 1.019 \\
 1.75 & 0.015 & 0.009 & 0.009  & 1.873 & 0.972 & 0.95  & 0.698 \\
 2. & 0.008 & 0.006  & 0.006  & 0.82 & 0.543 & 0.547  & 0.458 \\
 2.25 & 0.004 & 0.004 & 0.004  & 0.372 & 0.296 & 0.301  & 0.272 \\
 2.5 & 0.002 & 0.002 & 0.002  & 0.139 & 0.127 & 0.128  & 0.122 \\
 2.75 & 0 & 0  & 0 & 0 & 0  & 0 & 0 \\
 \hline
\end{tabular}
\hfill
\begin{tabular}{|c|ccc|cccc|}
\hline\multicolumn{8}{|c|}{$E_7$ Adjoint} \\ \hline
$n_f$ &  $a_ 2^*$ & $a_ 3^*$ & $a_ 4^*$ & $\gamma _ 2^*$ & $\gamma_ 3^*$ & $\gamma _ 4^*$ & $\gamma _ {\text{AO}}^*$\\ 
 1.5 & 0.02 & 0.009 & 0.009 & 5.374 & 1.924 & 1.854 & 1.019 \\
 1.75 & 0.01 & 0.006 & 0.006  & 1.873 & 0.972 & 0.956  & 0.698 \\
 2. & 0.006 & 0.004 & 0.004  & 0.82 & 0.543 & 0.548  & 0.458 \\
 2.25 & 0.003 & 0.002 & 0.003  & 0.372 & 0.296  & 0.301 & 0.272 \\
 2.5 & 0.001 & 0.001  & 0.001  & 0.139 & 0.127  & 0.128  & 0.122 \\
 2.75 & 0 & 0  & 0 & 0 & 0  & 0 & 0 \\
\hline
\end{tabular}\\
\vspace{5mm}
\begin{tabular}{|c|ccc|cccc|}
\hline\multicolumn{8}{|c|}{$E_8$ Adjoint/Fundamental} \\ \hline
$n_f$ &  $a_ 2^*$ & $a_ 3^*$ & $a_ 4^*$ & $\gamma _ 2^*$ & $\gamma_ 3^*$ & $\gamma _ 4^*$ & $\gamma _ {\text{AO}}^*$\\ 
 1.5 & 0.012 & 0.006 & 0.005  & 5.374 & 1.924 & 1.87  & 1.019 \\
 1.75 & 0.006 & 0.004 & 0.004  & 1.873 & 0.972 & 0.96  & 0.698 \\
 2. & 0.003 & 0.002 & 0.003  & 0.82 & 0.543  & 0.549  & 0.458 \\
 2.25 & 0.002 & 0.001  & 0.002  & 0.372 & 0.296  & 0.301 & 0.272 \\
 2.5 & 0.001 & 0.001  & 0.001  & 0.139 & 0.127  & 0.128 & 0.122 \\
 2.75 & 0 & 0  & 0 & 0 & 0  & 0 & 0 \\
 \hline
\end{tabular}
\caption{As in Table \ref{TBLExceptional} but for fermions in the adjoint representation.}
\end{table}

\begin{table}[ht!]
\begin{tabular}{|c|ccc|cccc|}
\hline\multicolumn{8}{|c|}{$SO(5)$} \\ \hline
$n_f$ &  $a_ 2^*$ & $a_ 3^*$ & $a_ 4^*$ & $\gamma _ 2^*$ & $\gamma_ 3^*$ & $\gamma _ 4^*$ & $\gamma _ {\text{AO}}^*$\\ 
 8.  & -2.833 & 0.12 & 0.126 & 722.156 & 2.097 & -1.924 & 0.841 \\
 9.  & 0.476 & 0.084 & 0.087  & 23.625 & 1.008 & -0.239 & 0.659 \\
 10.  & 0.188 & 0.062 & 0.066  & 4.404 & 0.612 & 0.107 & 0.514 \\
 11.  & 0.103 & 0.047 & 0.051  & 1.629 & 0.416 & 0.219 & 0.396 \\
 12.  & 0.062 & 0.035 & 0.038  & 0.765 & 0.296 & 0.234 & 0.297 \\
 13.  & 0.039 & 0.026 & 0.027  & 0.396 & 0.21 & 0.198  & 0.213 \\
 14.  & 0.023 & 0.017 & 0.018  & 0.206 & 0.1	4 & 0.141 & 0.141 \\
 15.  & 0.012 & 0.01 & 0.01  & 0.096 & 0.079 & 0.08  & 0.079 \\
 16.   & 0.003 & 0.003 & 0.003 & 0.026 & 0.025  & 0.025 & 0.025 \\
\hline
\end{tabular}
\hfill
\begin{tabular}{|c|ccc|cccc|}
\hline\multicolumn{8}{|c|}{$SO(8)$} \\ \hline
$n_f$ &  $a_ 2^*$ & $a_ 3^*$ & $a_ 4^*$ & $\gamma _ 2^*$ & $\gamma_ 3^*$ & $\gamma _ 4^*$ & $\gamma _ {\text{AO}}^*$\\ 
 8. & 0.472 & 0.051 & 0.049  & 127.899 & 2.557 & 0.787  & 1.016 \\
 9. & 0.128 & 0.036  & 0.037 & 11.005 & 1.312  & 0.516 & 0.797 \\
 10. & 0.066 & 0.027  & 0.029 & 3.458 & 0.808  & 0.451 & 0.622 \\
 11. & 0.039 & 0.021 & 0.023 & 1.542 & 0.546 & 0.408 & 0.478 \\
 12. & 0.025 & 0.016 & 0.017 & 0.797 & 0.383 & 0.344  & 0.359 \\
 13. & 0.016 & 0.011  & 0.012 & 0.437 & 0.266  & 0.263 & 0.258 \\
 14. & 0.01 & 0.008  & 0.008 & 0.237 & 0.175  & 0.177 & 0.171 \\
 15. & 0.005 & 0.004  & 0.004 & 0.114 & 0.097  & 0.099 & 0.096 \\
 16. & 0.001 & 0.001  & 0.001 & 0.031 & 0.03 & 0.03 & 0.03 \\
\hline
\end{tabular} \\
\vspace{5mm}
\begin{tabular}{|c|ccc|cccc|}
\hline\multicolumn{8}{|c|}{$SO(10)$} \\ \hline
$n_f$ &  $a_ 2^*$ & $a_ 3^*$ & $a_ 4^*$ & $\gamma _ 2^*$ & $\gamma_ 3^*$ & $\gamma _ 4^*$ & $\gamma _ {\text{AO}}^*$\\ 
 5. & 0.485 & 0.041 & 0.038 & 292.562 & 4.346  & 2.998 & 1.26 \\
 6. & 0.072 & 0.025 & 0.026  & 8.171 & 1.507 & 1.012  & 0.875 \\
 7. & 0.032 & 0.017 & 0.018  & 2.101 & 0.76 & 0.616  & 0.6 \\
 8. & 0.016 & 0.011 & 0.012  & 0.809 & 0.436 & 0.417  & 0.394 \\
 9. & 0.008 & 0.007 & 0.007  & 0.342 & 0.245 & 0.249  & 0.233 \\
 10. & 0.003 & 0.003 & 0.003  & 0.123 & 0.108 & 0.109  & 0.105 \\
 11. & 0 & 0 & 0 & 0 & 0  & 0 & 0 \\
\hline
\end{tabular}
\hfill
\begin{tabular}{|c|ccc|cccc|}
\hline\multicolumn{8}{|c|}{$SO(12)$} \\ \hline
$n_f$ &  $a_ 2^*$ & $a_ 3^*$ & $a_ 4^*$ & $\gamma _ 2^*$ & $\gamma_ 3^*$ & $\gamma _ 4^*$ & $\gamma _ {\text{AO}}^*$\\ 
 3. & 0.33 & 0.033  & 0.03 & 256.867 & 6.153 & 5.796 & 1.458 \\
 4. & 0.033 & 0.016  & 0.016 & 3.845 & 1.254 & 1.098 & 0.812 \\
 5. & 0.012 & 0.008 & 0.009  & 0.815 & 0.484 & 0.48  & 0.423 \\
 6. & 0.004 & 0.003  & 0.003 & 0.205 & 0.172  & 0.175 & 0.165 \\
\hline
\end{tabular}
\caption{As in Table \ref{TBLExceptional} but for fermions in the spinorial representation of some $SO(N)$ gauge groups.
\label{TBLSpin}}
\end{table}

\FloatBarrier

\end{document}